\documentclass[aps,prb,epsf,twocolumn,showpacs]{revtex4-1}

\usepackage{amsmath,amssymb,graphics,epsfig,epstopdf,color,verbatim,ulem,braket,tabularx,float}
\usepackage[colorlinks,linkcolor=blue,citecolor=blue,urlcolor=blue]{hyperref}

\begin{document}

\title{\mbox{Exact diagonalization study of the anisotropic Heisenberg model related to YbMgGaO$_4$}}

\author{Muwei Wu}
\author{Dao-Xin Yao}
\email{yaodaox@mail.sysu.edu.cn}
\author{Han-Qing Wu}
\email{wuhanq3@mail.sysu.edu.cn}
\affiliation{\mbox{Center for Neutron Science and Technology, School of Physics, Sun Yat-sen University, Guangzhou, 510275, China} \mbox{and State Key Laboratory of Optoelectronic Materials and Technologies, School of Physics, Sun Yat-sen University,} Guangzhou, 510275, China}

\begin{abstract}
  Employing exact diagonalization, we systematically study the anisotropic Heisenberg model which is related to rare-earth triangular-lattice materials. We probe its full 3D phase diagram afresh and identify a large region of quantum spin liquid (QSL) phase which can extend to the QSL region of the $J_1$--$J_2$ triangular Heisenberg model. Furthermore, we explore the magnetization curves of different phases and reproduce the 1/3-magnetization plateau in the quantum spin liquid phase region. More importantly, to study the possible chemical disorders in real materials, we consider the randomness of exchange interactions and find no spin glass order. And there is a large region of random-singlet phase which contains strongly random spin networks, dominated by two-spin singlets, four-spin singlets and other singlet domains. Our comprehensive ED study can give detailed insightful understanding of the microscopic Hamiltonian related to the YbMgGaO$_4$ and some other related rare-earth triangular-lattice materials.
\end{abstract}


\date{\today}
\maketitle

\section{Introduction}

Quantum spin liquid (QSL) phase~\cite{Wen1991, *Balents2010, *XGWen2002, *Kitaev2006, *Savary2016, *Norman2016,*ZYi2017, *Broholm2020} is an exotic quantum phase of matter beyond the Landau-Ginzburg-Wilson symmetry-breaking paradigm and displays rich physics, like nonlocal fractional excitations, long-range entanglement and emergent gauge field. QSLs are more likely to be found in frustrated spin systems, such as triangular and Kagome lattices. The geometric frustration and quantum fluctuation may prevent any magnetic long-range ordering even at zero temperature.

In recent years, two-dimensional rare-earth-based frustrated magnets play an important role and gain considerable efforts to realize the QSL phase. Among that, YbMgGaO$_4$~\cite{Li2015tri, Li2015tri2, Li2016tri, SYLi2016, Li2017tri, Li2017tri2, Shen2016, Paddison2017,YLi2019, Lima_2019, MMajumder2020, YLi2019II} and rare-earth chalcogenide family NaYbCh$_2$(Ch = O, S, Se)~\cite{WWLiu2018}, are perfect triangular layer compounds with no structural or magnetic transition down to very low temperature. Especially, the broad continuum of magnetic excitation in the inelastic neutron scattering reveals a possible $U(1)$ QSL with a spinon Fermi surface~\cite{Shen2016,ShenYao2018,PLDai2020}. Unprecedentedly, the magnetic excitation in the fully polarized state at sufficient high field remains very broad in both energy and wave vector, indicating the possible of disorders caused by the site-mixing of Mg/Ga, giving rising to the distributions of the effective spin-1/2 $g$ factors and the magnetic couplings~\cite{Li2017tri}. In fact, one recent experiment has observed some spin-glass-like behaviors both in the YbMgGaO$_4$ and its sister compound YbZnGaO$_4$~\cite{ZMa2017}. But other experiments seem exclude a true spin freezing in YbMgGaO$_4$~\cite{YLi2019,LShu2020}.

To understand macroscopic behaviors of these materials, an easy-plane XXZ Hamiltonian with anisotropic exchange interactions was proposed to describe the effective spin-1/2 interactions~\cite{Li2015tri2}. This microscopic Hamiltonian was studied by various numerical and analytical approaches~\cite{YDLi2016, CLLiu2016, YDLi2017, YDLi2017II, QLuo2017, ZYZhu2017, ZYZhu2018, LBalents2018, Jason2018, Maksimov2019, HQWu2019, SZLi2020}. In this paper, we use exact diagonalization (ED) to study this anisotropic Heisenberg model afresh. We depict the comprehensive 3D phase boundaries using extensive finite-size scaling. We have further studied the magnetic field effect and most importantly the bond randomness effect. The random-singlet phase under bond randomness in the model we studied had not been revealed before and will provide insightful understanding of the YbMgGaO$_4$ and other related materials.


\section{Model and Method}

The generic spin Hamiltonian of YbMgGaO$_4$ with the next-nearest-neighbor exchange interaction on the triangular lattice reads~\cite{Li2015tri2, Paddison2017},
\begin{equation*}
\begin{split}\begin{array}{l}
H = \sum\limits_{\left\langle {i,j} \right\rangle } {\left[ J_1{S_i^zS_j^z + \frac{\alpha J_1}{2} \left( {S_i^ + S_j^ -  + S_i^ - S_j^ + } \right)} \right.} \\
+ J_1^{ \pm  \pm } ({\gamma _{ij}}S_i^ + S_j^ +  + \gamma _{ij}^*S_i^ - S_j^ - )\\
\left. { - \frac{{iJ_1^{z \pm } }}{2}(\gamma _{ij}^*S_i^ + S_j^z - {\gamma _{ij}}S_i^ - S_j^z + \braket{i\leftrightarrow j})} \right]\\
 + \sum\limits_{\left\langle {\left\langle {i,j} \right\rangle } \right\rangle } {\left[ {J_2 S_i^zS_j^z + \frac{\alpha J_2}{2}\left( {S_i^ + S_j^ -  + S_i^ - S_j^ + } \right)} \right]} \\
+ \mu_0\mu_B\sum\limits_{i}{\left[g_{\perp} (h_x S_i^x + h_y S_i^y) + g_{\parallel} h_z S_i^z \right]},
\end{array}
\label{Eq:Hmlt}
\end{split}
\end{equation*}
where $J_1^{\pm\pm}$ and $J_1^{z\pm}$ arise from the strong spin-orbital coupling, $\gamma_{ij}=1, e^{-i2\pi/3}, e^{i2\pi/3}$ are for the bond along three principle axes, respectively. In the following calculations, we set the XXZ anisotropic $\alpha=1.317$~\cite{YDLi2018} and set $J_1=1$ as the energy unit. In the bond randomness case, the interaction strengths $J_{ij}$ are uniformly distributed in the range $\left[J_{ij}(1-\Delta),J_{ij}(1+\Delta)\right]$ which are controlled by $\Delta$. $\Delta =1$ corresponds to the strongest bond randomness case. In the following, we define $H_{\perp}=\mu_0\mu_B g_{\perp}\sqrt{h_x^2+h_y^2}, H_{\parallel}=\mu_0\mu_B g_{\parallel}h_z$ as the magnetic-field strengths to simplify the notations.

To get the phase boundaries, we have defined two kinds of magnetic order parameters. The first is the square sublattice magnetization for the $120^{0}$ N\'eel phase~\cite{TSakai2014, Kawamura2015, HQWu2019},
\begin{equation*}
m_{N}^{2}=\frac{1}{3}\sum_{\alpha=1}^{3}\left[\frac{1}{(N/6)(N/6+1)}\left\langle \left(\sum_{i\in\alpha}\mathbf{S}_{i}\right)^2\right\rangle\right],
\end{equation*}
where $\alpha=1,2,3$ represent the three sublattices of the $120^{0}$ order. The second is the square sublattice magnetization for the stripe phases~\cite{Kawamura2015, HQWu2019},
\begin{equation*}
m_{str}^{2}=\frac{1}{6}\sum_{v=1}^{3}\sum_{\beta_v=1}^{2}\left[\frac{1}{(N/4)(N/4+1)}\left\langle \left(\sum_{i\in\beta_v}\mathbf{S}_{i}\right)^2\right\rangle\right],
\end{equation*}
where $v=1,2,3$ represent three kinds of stripe orders, and $\beta_v=1,2$ represent the two sublattices of $v$-kind stripe order. We use the leading linear scaling $1/\sqrt{N}$ to estimate the magnetic orders in the thermodynamic limit. The finite-size clusters used in the ED calculations are shown in Appendix~\ref{App:FSClusters}.

To eliminate other conventional orders in the quantum spin liquid region, we have also calculated three kinds of structure factors. The first one is the chiral structure factor
\begin{equation*}
\chi(\mathbf{q})=\frac{1}{N}\sum\limits_{ij}e^{-i\mathbf{q}\mathbf{r}_{ij}}\langle\hat{\chi}_{i}\hat{\chi}_{j}\rangle,
  \hat{\chi_{i}}=\hat{\mathbf{S}}_{i}\cdot(\hat{\mathbf{S}}_{i+\mathbf{a}_{1}}\times\hat{\mathbf{S}}_{i+\mathbf{a}_{2}}),
\end{equation*}
where $\mathbf{a}_1=(a,0)$ and $\mathbf{a}_2=(a/2,\sqrt{3}a/2)$ are the primitive vectors of triangular lattice and we set the lattice constant $a=1$ as the unit length. The second one is the dimer structure factor
\begin{eqnarray*}
D(\mathbf{q}) &=& \frac{1}{3N}\sum\limits_{ij}\sum\limits_{pq}e^{-i\mathbf{q}\mathbf{r}_{ip,jq}}\left[\left\langle\hat{\mathbf{B}}_{ip}\hat{\mathbf{B}}_{jq}\right\rangle\right],\\
\hat{\mathbf{B}}_{ip} &=& \hat{\mathbf{S}}_{i}\hat{\mathbf{S}}_{i+p}-\langle\hat{\mathbf{S}}_{i}\hat{\mathbf{S}}_{i+p}\rangle,
\end{eqnarray*}
where $i+p$ means the nearest-neighbor site of $i$-site along $\mathbf{a}_{1}, \mathbf{a}_{2}, -\mathbf{a}_{1}+\mathbf{a}_{2}$ direction for $p=1,2,3$, respectively. $\mathbf{r}_{ip,jq}$ means the displacement between centers of two bonds. The third one is the spin freezing order parameter
\begin{equation*}
q=\frac{1}{N}\sqrt{\sum_{ij} \langle\hat{\mathbf{S}}_{i}\hat{\mathbf{S}}_{j}\rangle^2},
\end{equation*}
which is used to detect the possible spin-glass ordering.

\begin{figure}[ht!]
  \centering
  \includegraphics[width=0.45\textwidth]{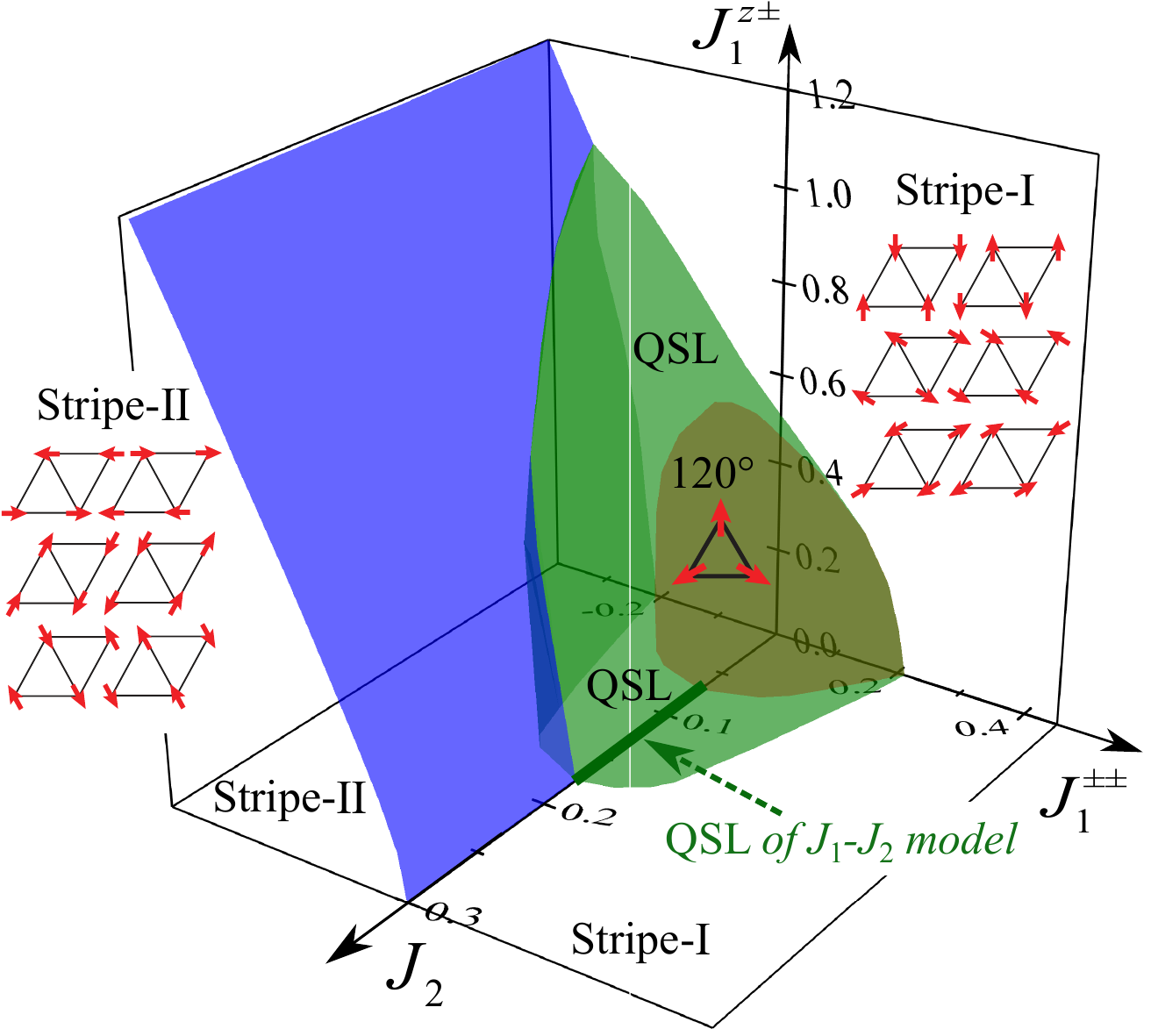}
  \caption{The 3D phase diagram of anisotropic triangular Heisenberg model related to YbMgGaO$_4$ in the $\alpha=1.317, J_2-J_1^{\pm\pm}-J_1^{z\pm}$ parameter space.  Four distinct phases, including 120$^{\circ}$ N\'eel phase, two stripe phases and a quantum spin liquid phase, are reproduced by our ED calculations. The magnetic structures of three magnetic ordered phases in the $xy$ plane are shown inside the phase regions, and the nonzero $J_1^{z\pm}$ will tilt the spins out of $xy$ plane.}
  \label{fig:PhaseDiagramI}
\end{figure}

\begin{figure}[ht!]
  \centering
  \includegraphics[width=0.515\textwidth]{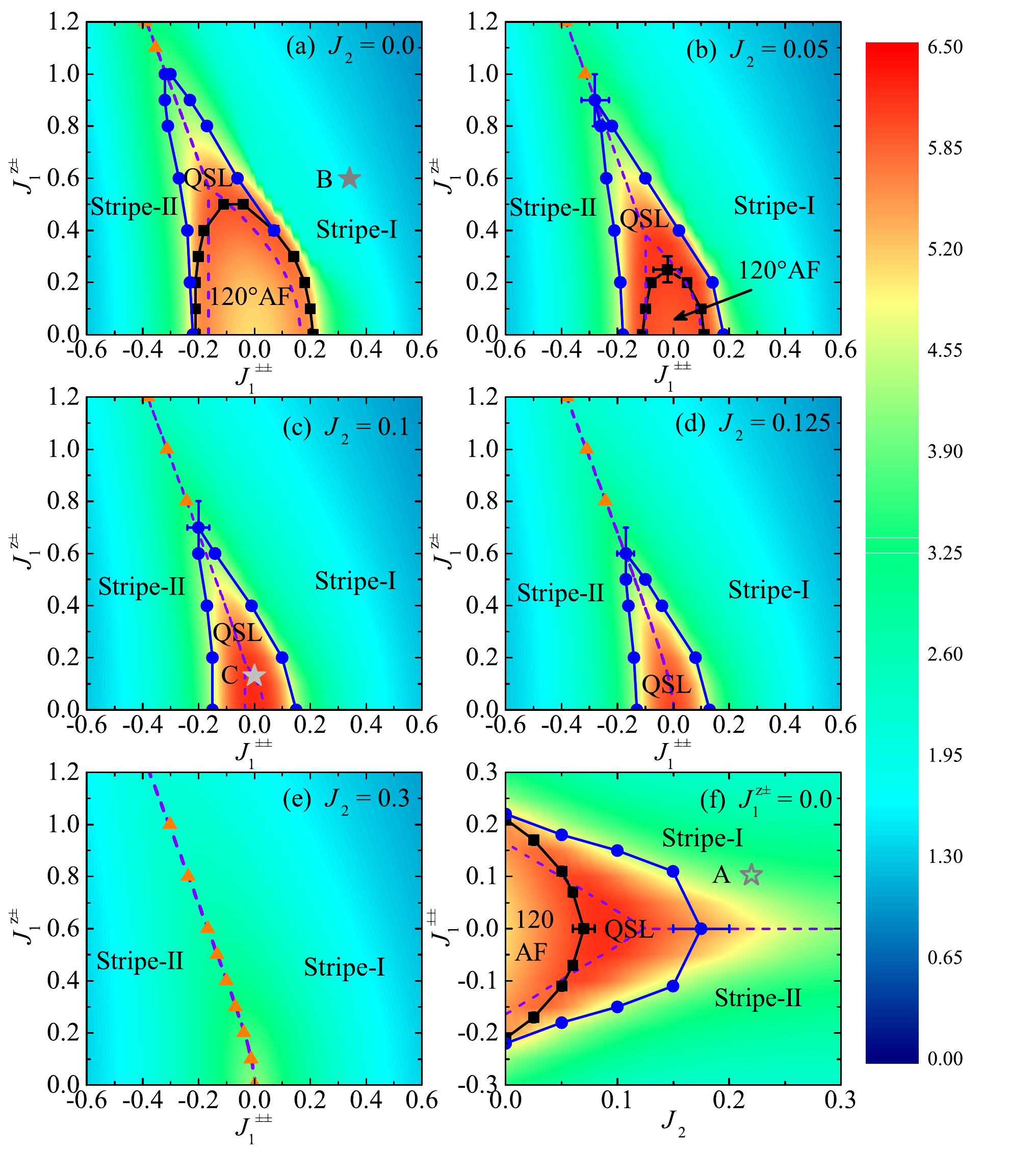}
  \caption{Phase diagrams on the slices of (a) $J_2=0, J_1^{\pm\pm}-J_1^{z\pm}$, (b)$J_2=0.05, J_1^{\pm\pm}-J_1^{z\pm}$, (c) $J_2=0.1, J_1^{\pm\pm}-J_1^{z\pm}$, (d)$J_2=0.125, J_1^{\pm\pm}-J_1^{z\pm}$, (e)$J_2=0.3, J_1^{\pm\pm}-J_1^{z\pm}$, and (f) $J_1^{z\pm}=0, J_2-J_1^{\pm\pm}$. The color bar shows the strength of the frustration parameter $f$ obtained by full exact diagonalization using 12-site cluster. The black and blue phase transition points are obtained from linear extrapolations of finite-size magnetic order parameters, while the yellow points are obtained by the level crossings of low excited energy states (see Appendix~\ref{App:Stripe}). The purple dashed lines are the classical phase transition lines between three magnetic phases. The star points A, B, and C shown in (f), (a), and (c) denote some sets of exchange parameters fitted by experimental data and got from Ref.~\onlinecite{Paddison2017}, Ref.~\onlinecite{YDLi2018}, and Ref.~\onlinecite{Haravifard2020}, respectively, while the hollow star point A used a different easy-plane anisotropic $\alpha\approx1.73$. Some sets of exchange parameters are outlined in Ref.~\onlinecite{XinshuZang2018}.}
  \label{fig:PhaseDiagramII}
\end{figure}

\begin{figure}[ht!]
  \centering
  \includegraphics[width=0.48\textwidth]{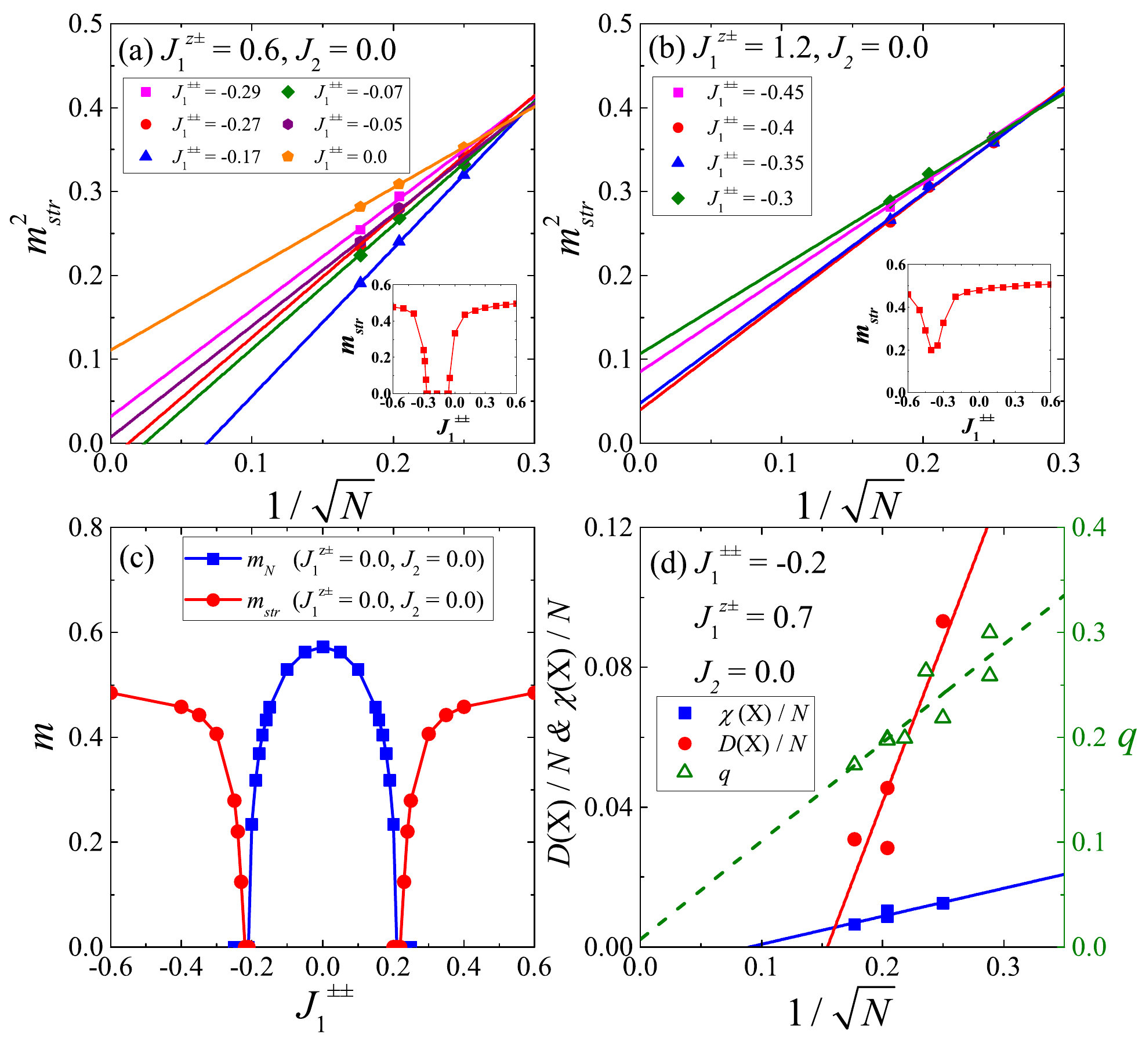}
  \caption{Linear extrapolations of the square sublattice magnetization for the stripe orders in selective paths which go along (a) $J_1^{z\pm}=0.6$ and (b) $J_1^{z\pm}=1.2$ horizontal lines in the phase slice of Fig.~\ref{fig:PhaseDiagramII}(a). Extrapolated stripe orders are shown in the insets. When $J_1^{z\pm}=0.6$, there are two phase transition points at around $J_{1c}^{\pm\pm}\approx-0.28$ and $J_{1c}^{\pm\pm}\approx-0.05$. While for $J_1^{z\pm}=1.2$, the extrapolated stripe order has a minimum at around $J_{1c}^{\pm\pm}\approx-0.39$ which is a signature of the first-order transition between two stripe phases. And the phase transition point is nearly the same as the classical one. (c) is the extrapolated magnetic orders along $J_1^{z\pm}=0$ horizontal lines. (d) shows the vanishing extrapolations of chiral, dimer and spin-freezing order parameters.}
  \label{fig:MagOrders}
\end{figure}

\section{Phase diagram}

\begin{figure}[htp!]
  \centering
  \includegraphics[width=0.48\textwidth]{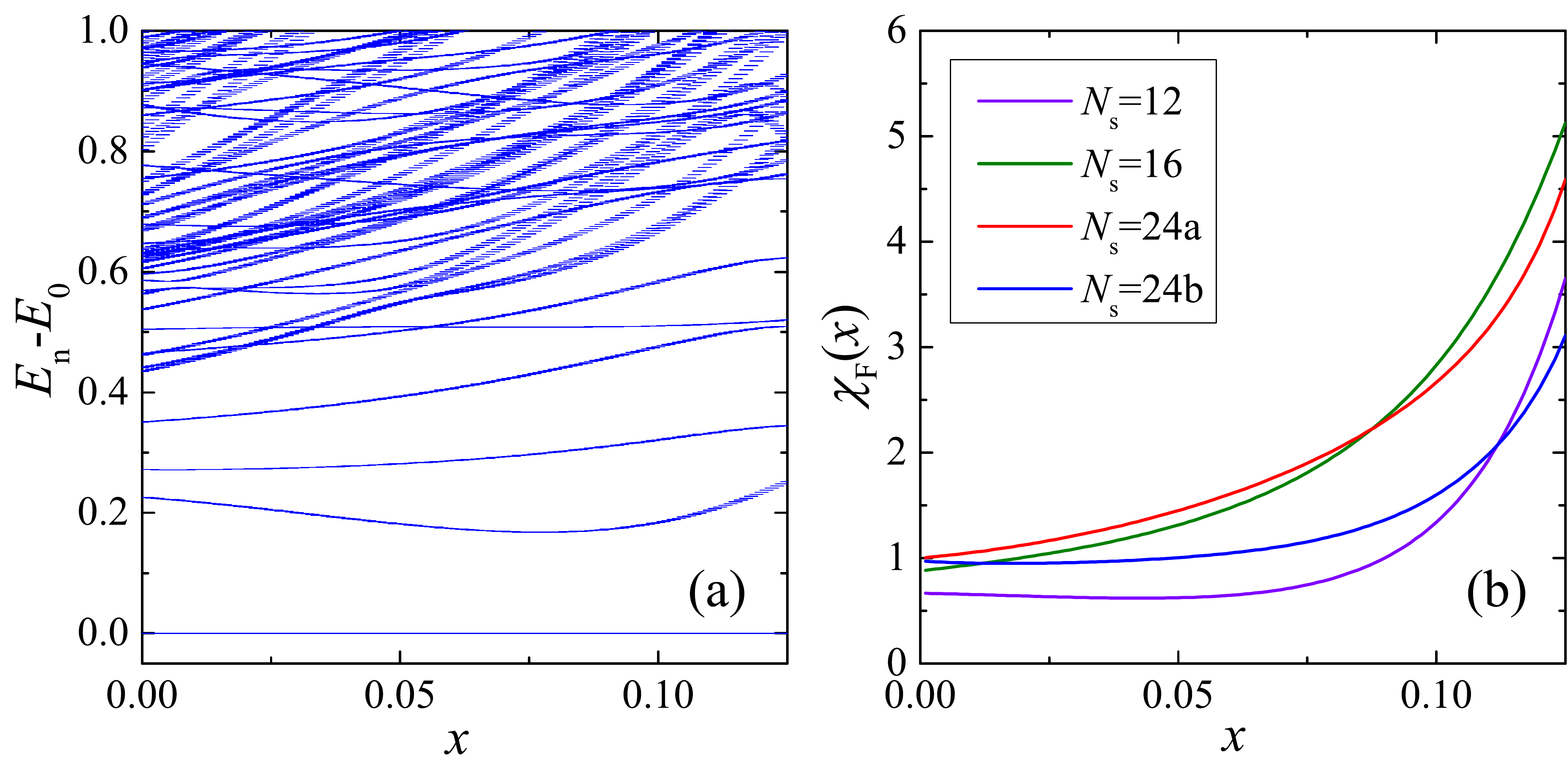}
  \caption{(a) Low-energy spectrum obtained by the $24a$ cluster and (b) fidelity susceptibility of different clusters change with control parameter $x$. Here, we take a straight-line path $J_2=x, J_1^{\pm\pm}=1.6x-0.2, J_1^{z\pm}=-5.6x+0.7, x\in[0,0.125]$ in the 3D parameter space to show the low-energy spectrum and the fidelity susceptibility. No level crossing or quasidegeneracy has been found by different clusters. And there is no any peak structure among $x\in [0.0,0.125]$ in the fidelity susceptibility. The increasing behavior near $x=0.125$ only indicates that it is close to the phase transition point between stripe phase and QSL.}
  \label{fig:QSLFD}
\end{figure}

The 3D phase diagram is illustrated in Fig.~\ref{fig:PhaseDiagramI}. Inside the dark yellow curve is the 120$^{\circ}$ N\'eel phase. The region in between the dark yellow curve and the green curve is a quantum spin liquid (QSL) phase. And the blue curve separates Stripe-I and Stripe-II phases. These stripe phases are Ising-like phases that have six degenerate ground states and a finite excitation gap according to the finite-size energy spectra (see Appendix~\ref{App:Stripe}). This degeneracy will be lifted after spontaneously $Z_6$ discrete symmetry breaking below a finite critical temperature in the thermodynamic limit~\cite{LBalents2018}. For the QSL, that region is a QSL based on two main reasons: one is that there are no conventional orders, including 120$^{\circ}$ N\'eel order, stripe order, dimer or valence-bond-solid order and spin-freezing order $q$ (see Fig.~\ref{fig:MagOrders}); another is that this phase can adiabatically connect to the QSL phase in the $J_1-J_2$ triangular Heisenberg model which can be identified by no any level crossing or avoided level crossing in the low-energy spectra and no any discontinuity or divergent tendency in the ground-state fidelity susceptibility $\chi_F(x)=\frac{2[1-F(x)]}{N(\delta x)^2}$ [see Fig.~\ref{fig:QSLFD} (b)], where the fidelity $F(x)=|\left\langle\Psi_0(x)|\Psi_0(x+\delta x)\right\rangle|$ measures the amounts of shared information between two quantum states. Meanwhile, we do not see any quasidegenerate states in the QSL region from our finite-size calculations [see Fig.~\ref{fig:QSLFD} (a)]. We conjecture that this QSL would be gapless in the thermodynamic limit similar to the $J_1-J_2$ triangular Heisenberg model~\cite{Imada2014, bishop2015, SRWhite2015, DNSheng2015, iqbal2016, ian2016, Jason2018,SJHu2019,Ferrari2019}. To give more details about this 3D phase diagram, we plot some slices in Fig.~\ref{fig:PhaseDiagramII}. The QSL boundaries are obtained by the vanishing of two kinds of magnetic orders: $120^{0}$ N\'eel order and stripe order. In Figs.~\ref{fig:MagOrders}(a)--\ref{fig:MagOrders}(c), we representatively show the linear extrapolations of magnetic orders along some horizontal paths on the $J_2=0, J_1^{\pm\pm}-J_1^{z\pm}$ slice. In addition, in Fig.~\ref{fig:PhaseDiagramII}, we use the contour plot to show the frustration parameter $f=|\Theta_{\rm{CW}}|/T_c$ on each slice, where $\Theta_{\rm{CW}}$ is the negative Curie-Weiss temperature and $T_c$ is the critical temperature. Here, we take the $T_c$ approximately as the temperature where the heat capacity gets its maximum value. Then we can confirm that the QSL region has a larger frustration parameter, especially after adding the next-nearest-neighbor $J_2$ interaction. The strong frustration in these regions prevents the magnetic ordering even at zero temperature. Under the guidance of the 3D phase diagram, we compare different sets of exchange parameters obtained by different research groups. Most of the parameter sets fall into the stripe phases. We only show three of them which is within or close to the QSL region, marked with A, B, and C in Fig.~\ref{fig:PhaseDiagramII}. Here we want to mention that the anisotropic exchange interactions $J_1^{\pm\pm}$ and $J_1^{z\pm}$ are weaker effects from the electron-spin resonance (ESR) measurements~\cite{Li2015tri2}. However, from our ED calculations, we find that the QSL region with only nearest-neighbor interactions needs a large $J_1^{z\pm}\sim 0.5 J_1$, but it would be reduced by adding the next-nearest-neighbor interaction or decreasing the XXZ anisotropic $\alpha$, which means $J_2$ is important to capture spin-liquid-behavior of triangular materials if one has to neglect the possible chemical disorders and let $\alpha\sim 1$. Compared with previous DMRG result from Ref.~\onlinecite{ZYZhu2018}, though different $\alpha$ is used, our QSL region in the $J_1^{\pm\pm}-J_1^{z\pm}, J_2=0$ plane is different to the DMRG one which is within the cone-like shape of 120$^{\circ}$ N\'eel ordered phase region. If we take a line path which connects two QSL regions in the $J_1^{\pm\pm}-J_1^{z\pm}, J_2=0$ and $J_2-J_1^{\pm\pm}, J_1^{z\pm}=0$ planes, we observe that the static spin structure factors $S(\mathbf{q})$ always have broad peaks at around M points, not at the K points observed in Ref.~\onlinecite{ZYZhu2018} with an inappropriate path which is inside the 120$^{\circ}$ N\'eel phase.

\section{Magnetic field effects}

\begin{figure}[t]
  \centering
  \includegraphics[width=0.49\textwidth]{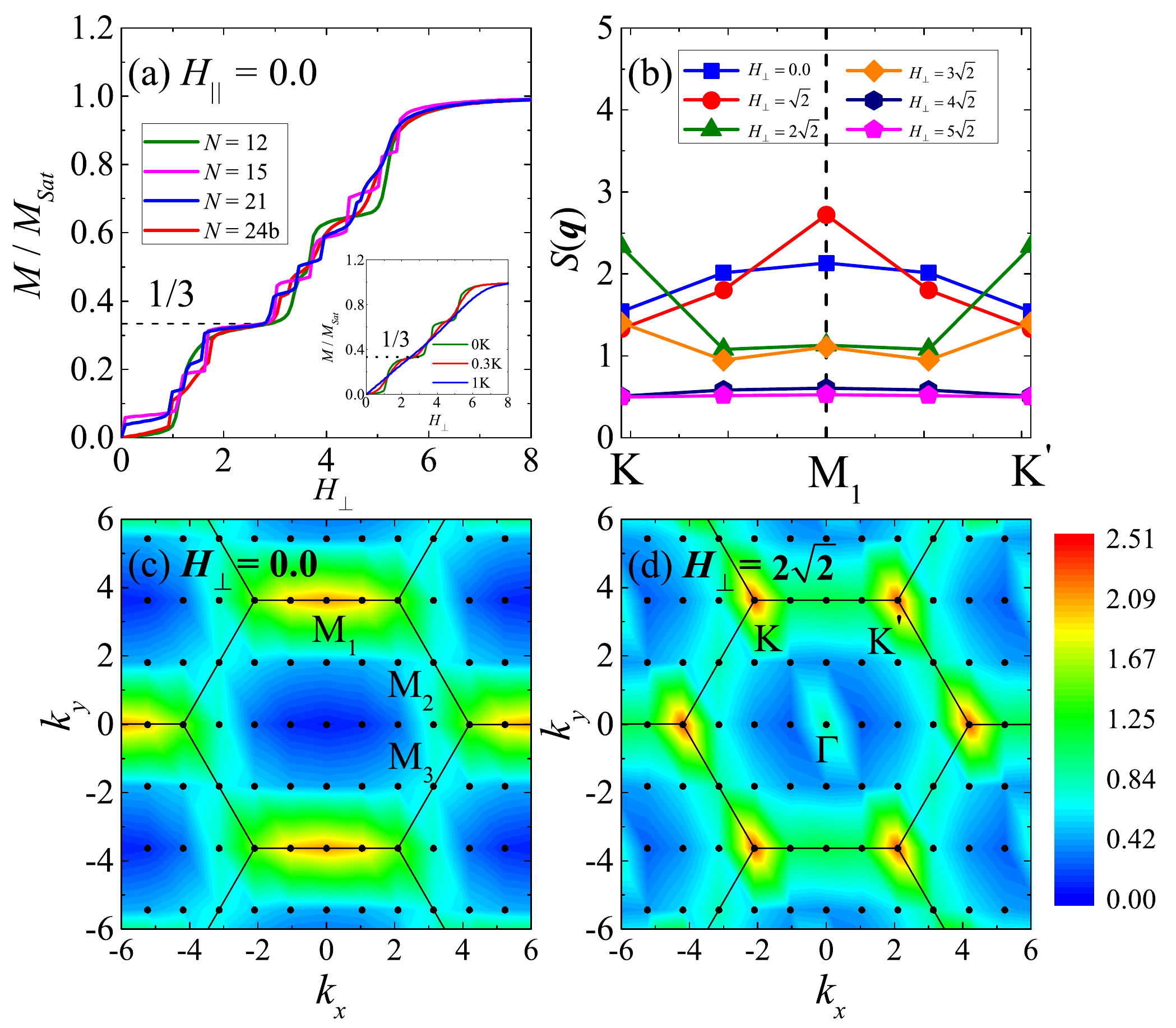}
  \caption{(a) Magnetization curves of the quantum spin liquid phase at $J_1^{\pm\pm}=-0.17, J_1^{z\pm}=0.6, J_2=0.0$ under external magnetic fields applied perpendicular to the $c$ axis. Combined the results of different clusters, a ``melting" 1/3 magnetization plateau is clearly shown near $H_{\perp}=2.5$. The inset shows the temperature dependence of magnetization curves obtained by 12-site cluster. (b) The spin structure factors $S(\mathbf{q})$ along $K\rightarrow M_1 \rightarrow K^{\prime}$ high symmetry path in the Brillouin zone (BZ) under different magnetic fields. With the increasing magnetic field, the spectral weight shifts from M points in the zero field to the K points around the plateau and then transfers to $\Gamma$ point in the fully polarized phase. Panels (c) and (d) are the contour plots of static spin structure factors in the whole BZ at $H_{\perp}=0$ and $2\sqrt{2}$, respectively. We use the $24b$ cluster to get those results in panels (b), (c), and (d).}
  \label{fig:MagFieldI}
\end{figure}

\begin{figure}[t]
  \centering
  \includegraphics[width=0.535\textwidth]{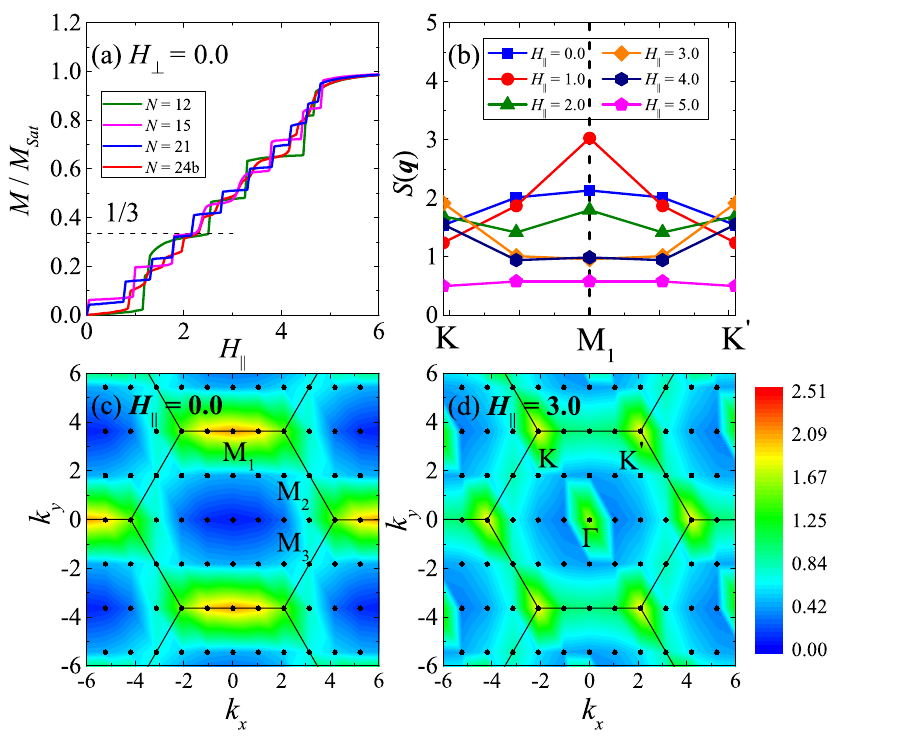}
  \caption{(a) Magnetization curves of the QSL phase at $J_1^{\pm\pm}=-0.17, J_1^{z\pm}=0.6, J_2=0.0$ under external magnetic field applied parallel to the $c$ axis. (b) The spin structure factors $S(\mathbf{q})$ along $K\rightarrow M_1 \rightarrow K^{\prime}$ high symmetry path in the Brillouin zone (BZ) under different strengths of magnetic field. Panels (c) and (d) are the contour plots of spin structure factors in the whole BZ at $H_{\parallel}=0$ and $3$, respectively. We use the $24b$ cluster to get those results in panels (b), (c), and (d). The intensity of $S(K)$ at $H_\parallel=3$ is weaker than the intensity of $S(M)$ at $H_\parallel=0$.}
  \label{fig:MagFieldII}
\end{figure}

We have studied the magnetization curves of three magnetic ordered phases and the quantum spin liquid phase. Here in the main text, we only show the magnetization curves at $J_1^{\pm\pm}=-0.17, J_1^{z\pm}=0.6, J_2=0.0$ of QSL region. When the magnetic field is applied perpendicular to the $c$ axis, though there is finite-size effect, we still can observe a clear ``melting" 1/3-magnetization plateau [see Fig.~\ref{fig:MagFieldI} (a)]. This 1/3-plateau or ``$uud$" phase is widely observed in the 120$^{\circ}$ N\'eel phase, but not in a QSL phase. The nonflatness of this plateau at zero temperature is due to the out of $xy$ plane anisotropic interaction $J_1^{z\pm}$. When the $J_1^{z\pm}$ further increases in the QSL region, this plateau melts to be a nonlinear rough curve. Another contribution to the nonflatness of the plateau without bond randomness is the temperature. When the temperature increases, the plateau will further melt to become a rough or even linear curve, which is shown in the inset of Fig.~\ref{fig:MagFieldI}(a). For the spin structure factor $S(\mathbf{q})$, we can observe that the spectral weight shifts from M points in the zero field to the K points in the sufficient strong field around the 1/3-magnetization plateau, and then transfers to the $\Gamma$ point in the fully polarized region. Interestingly, the recent experiment on the YbMgGaO$_4$~\cite{AA2020, Haravifard2020} with very low temperature has discovered the nonlinearity of the magnetization curve which may be a signature of the remnant of 1/3-magnetization plateau. The DMRG and classical Monte Carlo simulations~\cite{Haravifard2020} using the $C$ set of parameters [see Fig.~\ref{fig:PhaseDiagramII} (c)] have reproduced the nonlinearity of the magnetization curve. Here, our ED method has reproduced the similar behaviors not only in the $C$ set of parameters but also in the large region of QSL phase (see Appendix~\ref{App:Magnetization}). What's more, adding $J_2$ do not obviously change the flatness of the plateau, but the interval of the 1/3-plateau will shrink and disappear, while the 1/2-plateau will appear at larger $J_2$~\cite{Hiroki2017}. In addition, we have also studied the magnetization curves with the magnetic field parallel to the $c$ axis which can be seen in Fig.~\ref{fig:MagFieldII}. The 1/3-plateau seems still visible at $J_1^{\pm\pm}=-0.17, J_1^{z\pm}=0.6, J_2=0.0$, but has a quite narrow interval which is due to the easy-plane anisotropy $\alpha$ and the out of plane anisotropic interactions $J_1^{z\pm}$.

\section{Bond randomness effects}

\begin{figure}[t]
  \centering
  \includegraphics[width=0.49\textwidth]{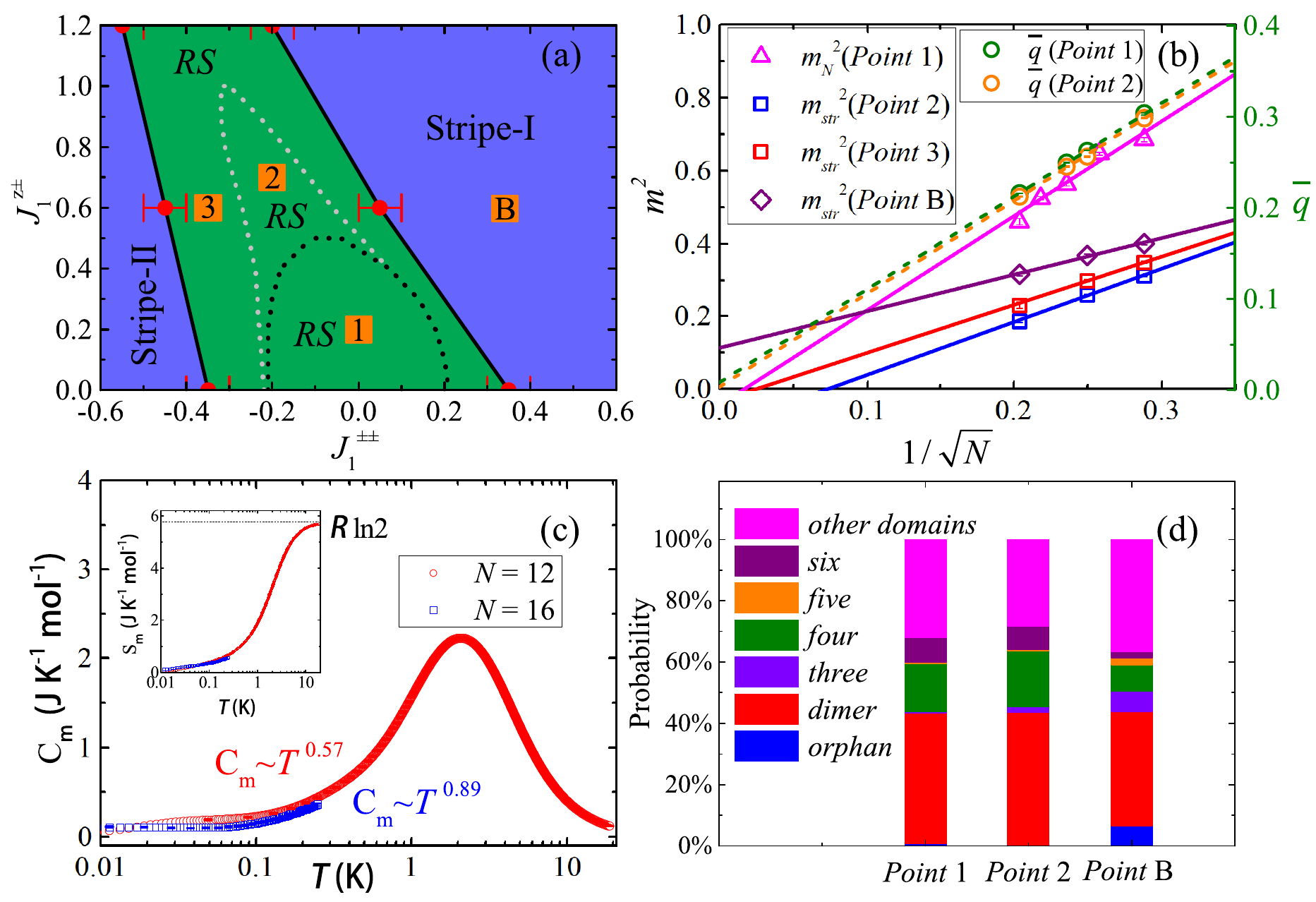}
  \caption{(a) Phase diagram under the strongest bond randomness $\Delta=1$ on the $J_2=0, J_1^{\pm\pm}-J_1^{z\pm}$ slice. The dashed lines are the phase boundaries in the clean $\Delta=0$ case. (b) Linear extrapolations of the magnetic orders and the average freezing parameter at different sets of parameters. At least 50 bond-randomness samples have been used to get the converged average values. (c) Magnetic heat capacities $C_m$ at $Point$ B obtained by 12 and 16 clusters. We used at least 200 bond-randomness samples to get the converged $C_m$ here. The inset shows the magnetic entropy $S_m=S_0+\int_0^T \frac{C_m}{T}dT$. No residual entropy (i.e., $S_0$=0) is found at low temperature. (d) Histogram of spin domains with different number of spins obtained by 150 independent random samples.}
  \label{fig:GlassOrder}
\end{figure}

To study the possible chemical disorders in real materials, like Ga/Mg mixing in YbMgGaO$_4$, we add uniform bond randomness into the Hamiltonian. Other distributions of the random exchange interactions do not change the conclusion qualitatively. For two stripe phases with finite excitation gaps, these magnetic orders deep into magnetic phases are stable to the bond randomness and persist up to the strongest randomness $\Delta=1$. Surprisingly, at B set of parameters, though the stripe order is still finite [see Fig.~\ref{fig:GlassOrder}(b)], the magnetic heat capacity $C_m$ under the strongest bond randomness [see Fig.~\ref{fig:GlassOrder}(c)] is similar to the experimental results~\cite{Li2015tri, Li2015tri2, SYLi2016, Paddison2017, ZMa2017}, other sets of parameters cannot reproduce the correct shape of the heat capacity. And the power-law exponents $\delta$ are $0.57$ and $0.89$ for 12 and 16 clusters, respectively. This power-law heat capacity is due to the nonzero density of low-lying excitations under bond randomness~\cite{Senthil2018,Kawamura2019}. For the $120^0$ N\'eel order, it is fragile to bond randomness, but it can persist up to a critical bond randomness strength $\Delta_c<1$~\cite{HQWu2019}. So in the strongest bond randomness $\Delta=1$ case, not only the QSL region and the entire $120^0$ phase region but also the stripe phase regions which are close to the phase boundaries [see the phase diagram in Figs.~\ref{fig:GlassOrder}(a) and \ref{fig:GlassOrder}(b)] will show nonmagnetic spin-liquid-like behavior. To detect the possible spin-glass order induced by the bond randomness, we show the average spin freezing parameters in Fig.~\ref{fig:GlassOrder}(b). They all are extrapolated to zero. There would be no spin-glass order even in the strongest bond randomness. This nonmagnetic spin-liquid-like phase in the $\Delta=1$ limit is actually a 2D analog of random-singlet (RS) phase~\cite{SkMa1979, DSFisher1994, Senthil2018,Kawamura2019}.

\begin{figure}[t]
  \centering
  \includegraphics[width=0.5\textwidth]{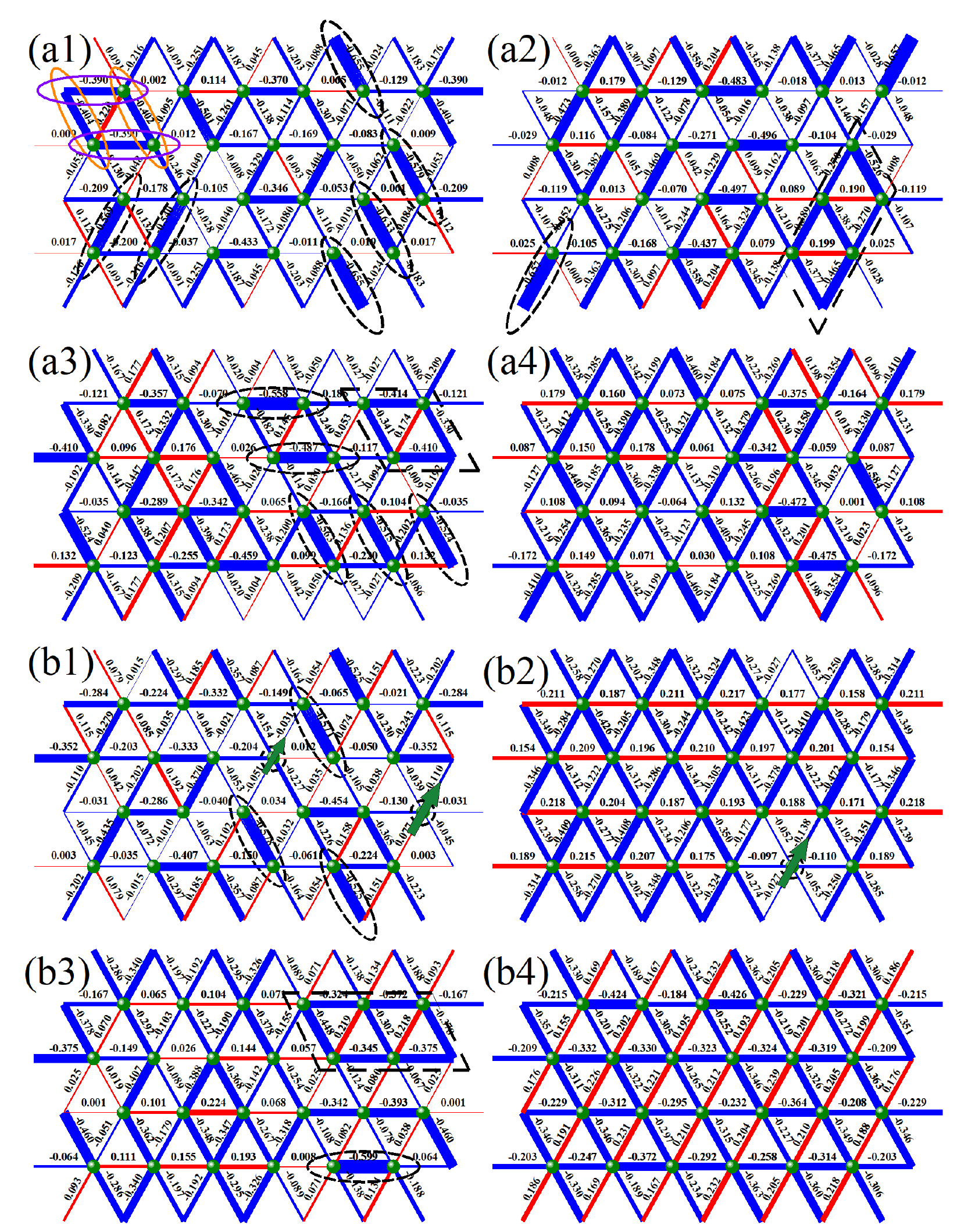}
  \caption{Nearest-neighbor spin correlations $\braket{\mathbf{S}_i\mathbf{S}_j}$ under specific bond-randomness configurations at [(a1)--(a4)] $Point$ 2 and [(b1)--(b4)] $Point$ B, respectively. Some distinct orphan spins, two-spin singlets and other singlet clusters are marked by dotted boxes.}
  \label{fig:BNNCorr}
\end{figure}

To describe this phase more clearly, we analyze the distribution of different spin domains under some different random exchange interaction configurations in Fig.~\ref{fig:GlassOrder}(d). In RS phase region, we find mostly the local two-spin singlets or dimers ($43\%$), four-spin singlets or resonating dimers ($16\%$ for $Point$ 1 and $18\%$ for $Point$ 2) and other larger singlet domains with even number of spins (resonating dimer domains). However, it is hard to find distinct orphan spins ($0.7\%$ for $Point$ 1 and $0.3\%$ for $Point$ 2) and long-distance two-spin singlets in RS phase region. In contrast, in stripe phase like the $B$ set of parameters, the fraction of orphan spins ($6\%$) becomes significant. Furthermore, the spin domains with odd numbers of spins also become nonnegligible. These domains with odd number of sites (including the orphan spins) will contribute to the Curie-law $\propto1/T$ of magnetic susceptibility at low temperature. More significantly, larger domains with stripe order contribute to the nonzero average magnetic order under the bond randomness. However, the finite temperature would further fragment the stripe ordered domains making it hard to be detected in experiment. In order to build an intuition of the random spin networks, in Fig.~\ref{fig:BNNCorr}, we representatively show the nearest-neighbor spin correlations $\braket{\mathbf{S}_i\mathbf{S}_j}$ under specific bond-randomness configurations at $Point$ 2 and $Point$ B, respectively. We can see the formations of orphan spins (marked by green arrows), two-spin singlets (marked by dotted oval box), four-spin singlets (resonating two types of singlet pairs marked by yellow and purple oval box) and other larger spin domains. Here we set the criterion of two spins belong to the same domain as the spin correlation between them is lower than $-0.25$ (see Ref.~\onlinecite{Kawamura2019} for more details). The singlet domains under bond randomness can help to clarify the continuous low-energy excitation and the absence of spinon contribution to thermal conductivity~\cite{ZMa2017,YLi2019,Li2017tri2}. And the last not the least, for the magnetization curve, the 1/3-magnetization plateau will further melt by the randomness of exchange interactions and $g$-factors~\cite{TSakai2014}, similar to the temperature effect. There would be no distinct 1/3-magnetization plateau under the strongest bond randomness~\cite{AA2020}.


\section{Summary and discussion}

In summary, we have used extensive finite-size scalings with ED to get the entire phase diagram in the 3D parameter space. Besides two gapped stripe phases and 120$^{\circ}$ N\'eel phase, there is a large QSL region extending to the QSL phase of the $J_1-J_2$ triangular Heisenberg model. After applying external magnetic fields, a 1/3-magnetization plateau can be observed at large region of QSL phase when the magnetic field is perpendicular to the $c$ axis. Most importantly, in the strongest bond randomness case, numerical result shows a large region of spin-liquid-like phase which is a 2D analog of random-singlet phase. It contains two-spin singlets, four-spin singlets and other larger spin singlets which have not been unveiled in the model we studied. In addition, our 3D phase diagram with different XXZ anisotropy (see Appendix~\ref{App:XXZ}) can also help to understand the QSL like behavior in AYbCh$_2$ (A=Na and Cs, Ch=O,S,Se)~\cite{WWLiu2018, PLDai2020, Doert2018, Ranjith2019, Ding2019, JXing2019, RSarkar2019, GuoJie2020, JMa2020, Wilson2020, Wilson2020II, Ranjith2019II, YTJia2020, ZHZhang2020, ZHZhang2020II, ZhZhang2021, Sichelschmidt_2019, Sichelschmidt2020, ZZangeneh2019}, Na$_2$BaCo(PO$_4$)$_2$~\cite{RZhong2019, NLi2020, SLee2021}.

Both YbMgGaO$_4$ and NaYbCh$_2$ share the same space symmetry group $R\bar{3}m$ and the perfect triangular magnetic layers which consist of Yb$^{3+}$ ions. Recent experiments~\cite{WWLiu2018,Doert2018,Ding2019,GuoJie2020,Wilson2020,Wilson2020II} have revealed that the interlayer Yb-Yb distance in NaYbO$_2$ and NaYbS$_2$ is shorter than YbMgGaO$_4$, which means the interlayer interactions may be relevant in the low temperature exchange model. A more complicated exchange Hamiltonian with the same in-plane part as the YbMgGaO$_4$ and compass-like interlayer exchange interactions has been proposed to understand macroscopic behaviors of these materials [see Ref.~\onlinecite{Wilson2020II} for more details]. In addition, we still need to caution about the possible randomness effects on NaYbCh$_2$, such as Na sites occupied by the Yb ions in NaYbSe$_2$~\cite{PLDai2020}. All of these issues, including the interlayer interactions and magnetic impurities between Yb layers, need further study.

\begin{acknowledgments}

\textit{Acknowledgments} H.Q.W. thanks Shou-Shu Gong, Wei Zhu, and Rong-Qiang He for helpful discussions. D.X.Y. is supported by NKRDPC-2018YFA0306001, NKRDPC-2017YFA0206203, NSFC-11974432, GBABRF-2019A1515011337, and Leading Talent Program of Guangdong Special Projects. H.Q.W. is supported by NSFC-11804401 and the Fundamental Research Funds for the Central Universities (Grant No. 19lgpy266).

\end{acknowledgments}

\appendix

\section{Finite-size clusters}
\label{App:FSClusters}

\begin{figure}[htp!]
  \centering
  \includegraphics[width=0.47\textwidth]{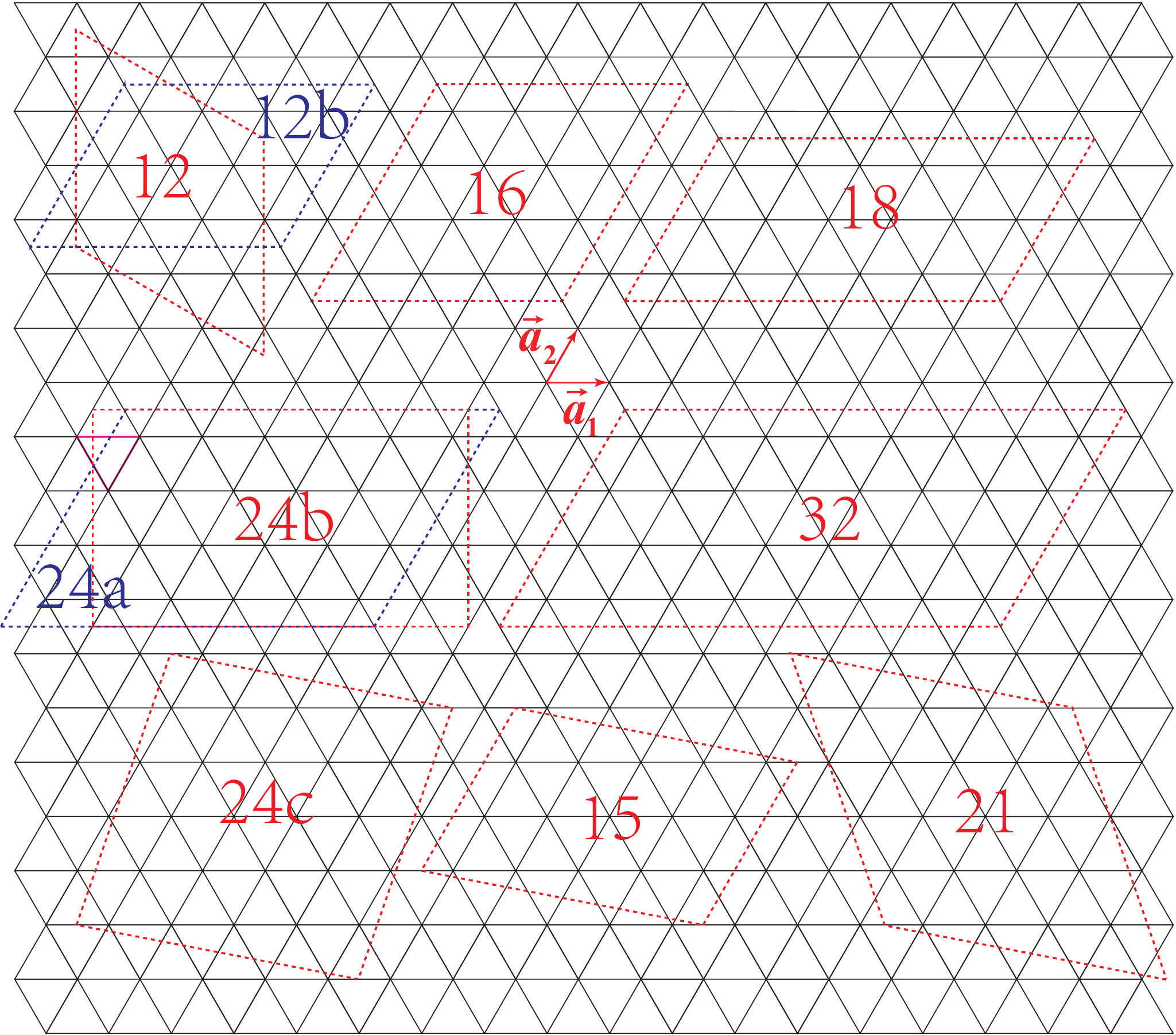}
  \caption{Finite-size clusters used in the ED calculations. $\mathbf{a}_{1}=(a,0)$ and $\mathbf{a}_{2}=(a/2,\sqrt{3}a/2)$ are primitive vectors of the triangular lattice. The $12$ cluster with $C_3$ symmetry has been used to do the full exact diagonalization and calculate the frustration parameter.}
  \label{fig:AppClusters}
\end{figure}

In this paper, we mainly use Lanczos exact diagonalization to get the 3D phase diagram and the low-energy spectrum. Meanwhile, we also employ full exact diagonalization to study the finite-temperature properties, such as heat capacity and magnetic susceptibility. To reduce the computational cost, we have used translation symmetry to do block diagonalization. The largest system size in the Lanczos calculations is 32 with the subspace of the largest block up to 0.13 billion.

Ten clusters are mainly used in our ED calculations which are shown in Fig.~\ref{fig:AppClusters}, denoted as $12$, $12b$, $15$, $16$, $18$, $21$, $24a$, $24b$, $24c$ and $32$, respectively. The $12$, $16$, $24a$, $24b$, and $32$ clusters have three $M$ momentum points which are significant for the stripe phases. These three-momentum points denote as $M_1=\frac{1}{2}b_2, M_2=\frac{1}{2}(b_1+b_2), M_3=\frac{1}{2}b_1$, where $b_1=(\frac{2\pi}{a},-\frac{2\pi}{\sqrt{3}a}),b_2=(0,\frac{4\pi}{\sqrt{3}a})$ are primitive lattice vectors in reciprocal space, $a=1$ is the lattice constant. Among these five clusters with even number of lattice site, the 12 and 24b clusters also contain two K points, $K_1=\frac{1}{3}b_1+\frac{2}{3}b_2, K_2=\frac{2}{3}b_1+\frac{1}{3}b_2$. The K points are important for 120$^{\circ}$ N\'eel phase and the 1/3-magnetization plateau phase or ``$uud$" phase. So we use the $12$, $15$, $18$, $21$, $24b$, and $24c$ clusters which contain K points to do the linear extrapolations of 120$^{\circ}$ N\'eel order and to study the 1/3-magnetization plateau. In the extrapolation of the spin freezing order parameters, we also use $12b$ cluster.

Here, we want to mention that three $M$ momentum points are nonequivalent in the $24a$, $24b$ and $32$ clusters which do not respect the $C_3$ rotation symmetry. Therefore, there may be only one $M$ point which has a broad peak in the spin structure factor $S(\mathbf{q})$ of QSL region [see Fig.~\ref{fig:MagFieldI}(c), Fig.~\ref{fig:MagFieldII}(c) in the main text, and Fig.~\ref{fig:AppMagStrFct}(b1)--\ref{fig:AppMagStrFct}(b4)]. We should see the diffuse magnetic scattering at around all three $M$ points when we use the clusters which have equivalent $M$ points, such as $16$ [see Fig.~\ref{fig:AppMagStrFct}(a1)--\ref{fig:AppMagStrFct}(a4)] and $36$ clusters.

\section{Conventional orders}
\label{App:Order}

\begin{figure}[htp!]
  \centering
  \includegraphics[width=0.49\textwidth]{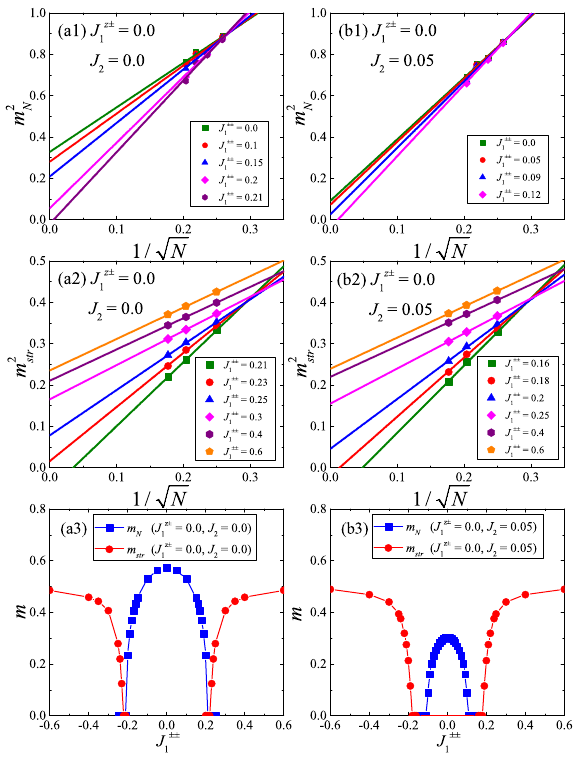}
  \caption{Linear extrapolations of the square sublattice magnetization for [(a1)--(b1)] the $120^{0}$ N\'eel phase and [(a2)--(b2)] the stripe phases. (a3) and (b3) are the extrapolated magnetic order parameters along $J_1^{z\pm}$ = 0 horizontal lines in Figs.~\ref{fig:PhaseDiagramII}(a) and ~\ref{fig:PhaseDiagramII}(b) of the main text, respectively. The $120^{0}$ N\'eel phase (blue, square) is sandwiched by two stripe phases (red, circle) at $J_2 = 0$. While at $J_2 = 0.05$, the QSL phase extends to the $J_1^{z\pm}$ = 0.0 region.}
  \label{fig:AppMagOrdersI}
\end{figure}

\begin{figure}[htp!]
  \centering
  \includegraphics[width=0.48\textwidth]{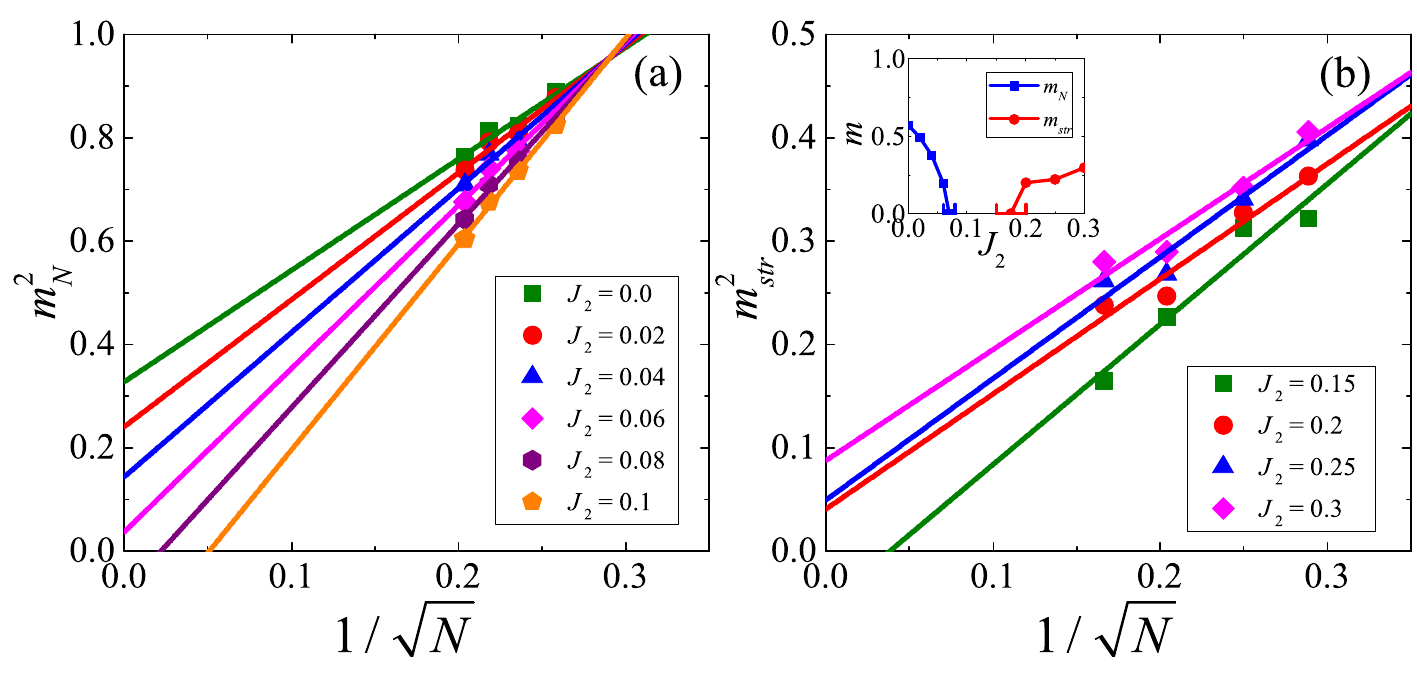}
  \caption{Linear extrapolations of the square sublattice magnetization for (a) the $120^{0}$ N\'eel phase and (b) the stripe phases along $J_1^{\pm\pm}$ = 0 horizontal lines in Fig.~\ref{fig:PhaseDiagramII}(f) of the main text. The inset of (b) is the extrapolated magnetic order parameters. The QSL phase is sandwiched between the $120^{0}$ N\'eel phase (blue, square) and the stripe phase (red, circle).}
  \label{fig:AppMagOrdersII}
\end{figure}

We have representatively shown the linear extrapolations of 120$^{\circ}$ N\'eel order and the stripe orders in the main text. Here, we want to show more details about the extrapolations, which can be seen in Figs.~\ref{fig:AppMagOrdersI} and \ref{fig:AppMagOrdersII}. The magnetic order parameters (square root of the extrapolated results) obtained from Figs.~\ref{fig:AppMagOrdersI}(a1) and \ref{fig:AppMagOrdersI}(a2), \ref{fig:AppMagOrdersI}(b1) and \ref{fig:AppMagOrdersI}(b2) are shown in Figs.~\ref{fig:AppMagOrdersI}(a3) and \ref{fig:AppMagOrdersI}(b3), respectively. Here, we mention that the stripe phases are Ising-like phases which have strong magnetic orders and weaker quantum fluctuations. Therefore, the linear extrapolations of magnetic orders are good enough to identify the phase boundaries. The extrapolated results will not change much when using some larger system sizes.

\begin{figure}[htp!]
  \centering
  \includegraphics[width=0.54\textwidth]{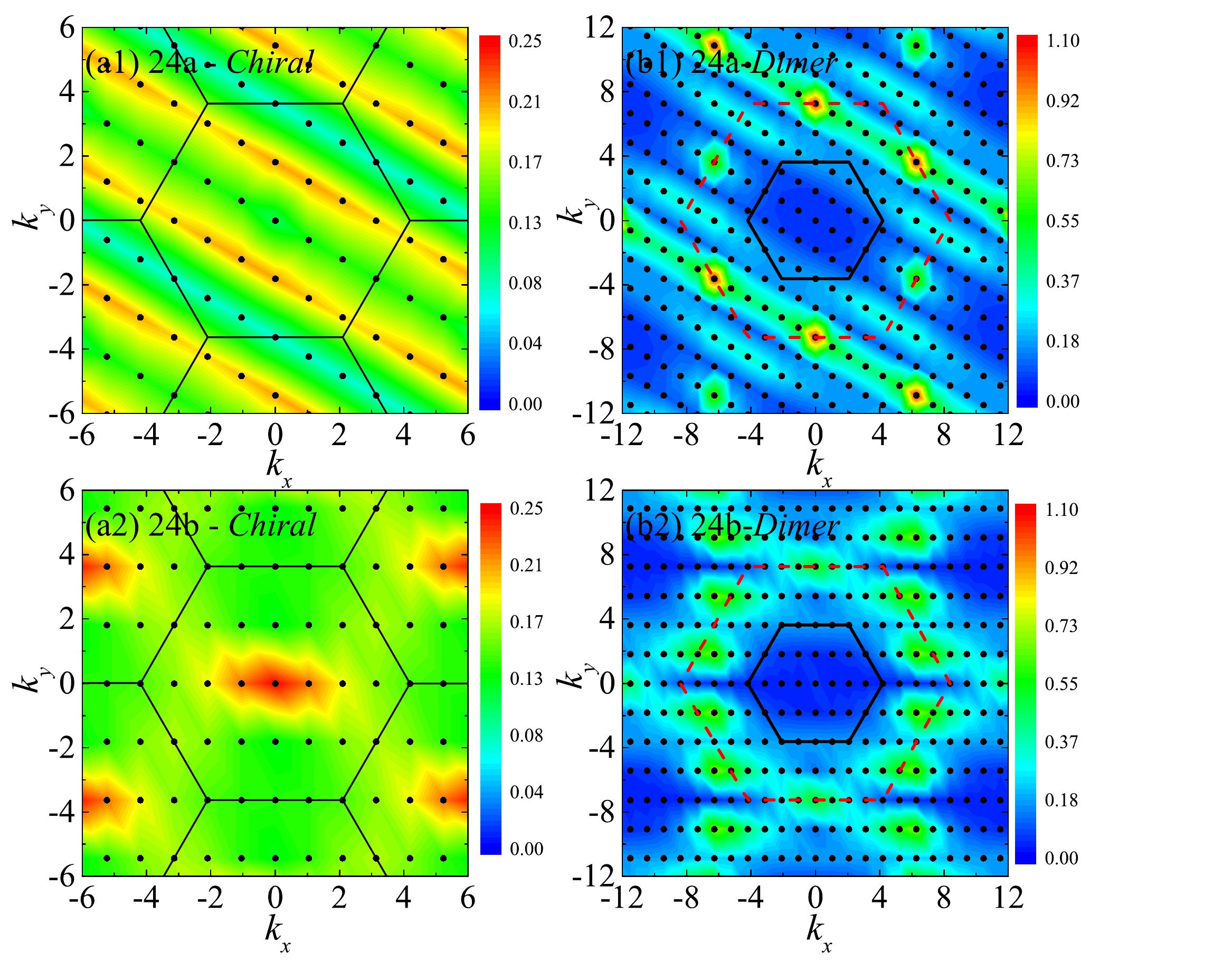}
  \caption{The contour plot of (a1, a2) chiral structure factor $\chi(\mathbf{q})$ and [(b1), (b2)] dimer structure factor $D(\mathbf{q})$ obtained by $24a$ and $24b$ cluster at $J_2=0, J_1^{\pm\pm}=-0.2, J_1^{z\pm}=0.7$. The black lines are the Brillouin zone edge of the original triangular lattice and the red dash line in [(b1), (b2)] is the Brillouin zone edge of the new kagome lattice.}
  \label{fig:AppChiral}
\end{figure}

To eliminate other conventional orders in the nonmagnetic phase region, we have also calculated the chiral and dimer structure factors. In our finite-size calculations, we find the peak positions of $\chi(\mathbf{q})$ and $D(\mathbf{q})$ vary between different clusters, which can be seen in Fig.~\ref{fig:AppChiral}. So we use $X$ to represent the wave vector where the peak is in Fig.~\ref{fig:MagOrders}(d) of the main text. We get the vanishing order parameters of these two conventional ordering using linear extrapolations. Therefore, the large region of nonmagnetic phase in the 3D parameter space (see Figs.~\ref{fig:PhaseDiagramI} and \ref{fig:PhaseDiagramII}) has no 120$^{\circ}$ N\'eel order, stripe orders, chiral order, dimer order and spin-freezing order [see Fig.~\ref{fig:MagOrders}(d)], and it is a quantum spin liquid phase.

\section{Stripe-I and Stripe-II phases}
\label{App:Stripe}

\begin{figure}[htp!]
  \centering
  \includegraphics[width=0.485\textwidth]{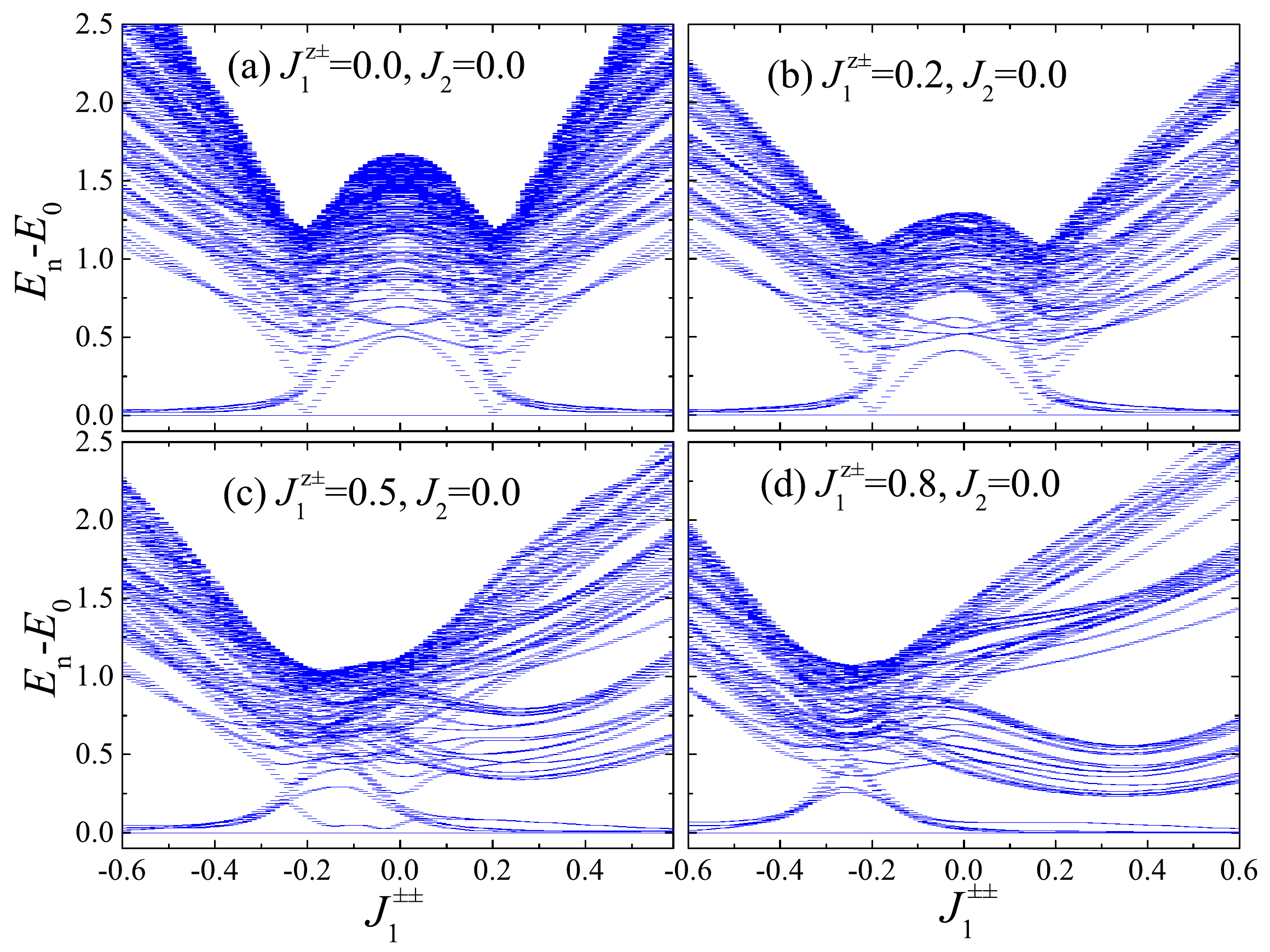}
  \caption{Low-energy spectra of $24a$ cluster with nearest-neighbor anisotropic interactions ($J_2=0$). There are six degenerate ground states and a finite excitation gap in the stripe-I and stripe-II phase regions. The ``inverse V-shape" in the low-energy spectra is more clear when $J_1^{z\pm}$ becomes larger. And the tip of ``inverse V-shape" can be used to identify the direct (first-order) phase transition points between two stripe phases on different slices [see the yellow triangular point in Fig.~\ref{fig:PhaseDiagramII} of the main text]. }
  \label{fig:AppStripeEngy}
\end{figure}

\begin{figure}[b]
  \centering
  \includegraphics[width=0.475\textwidth]{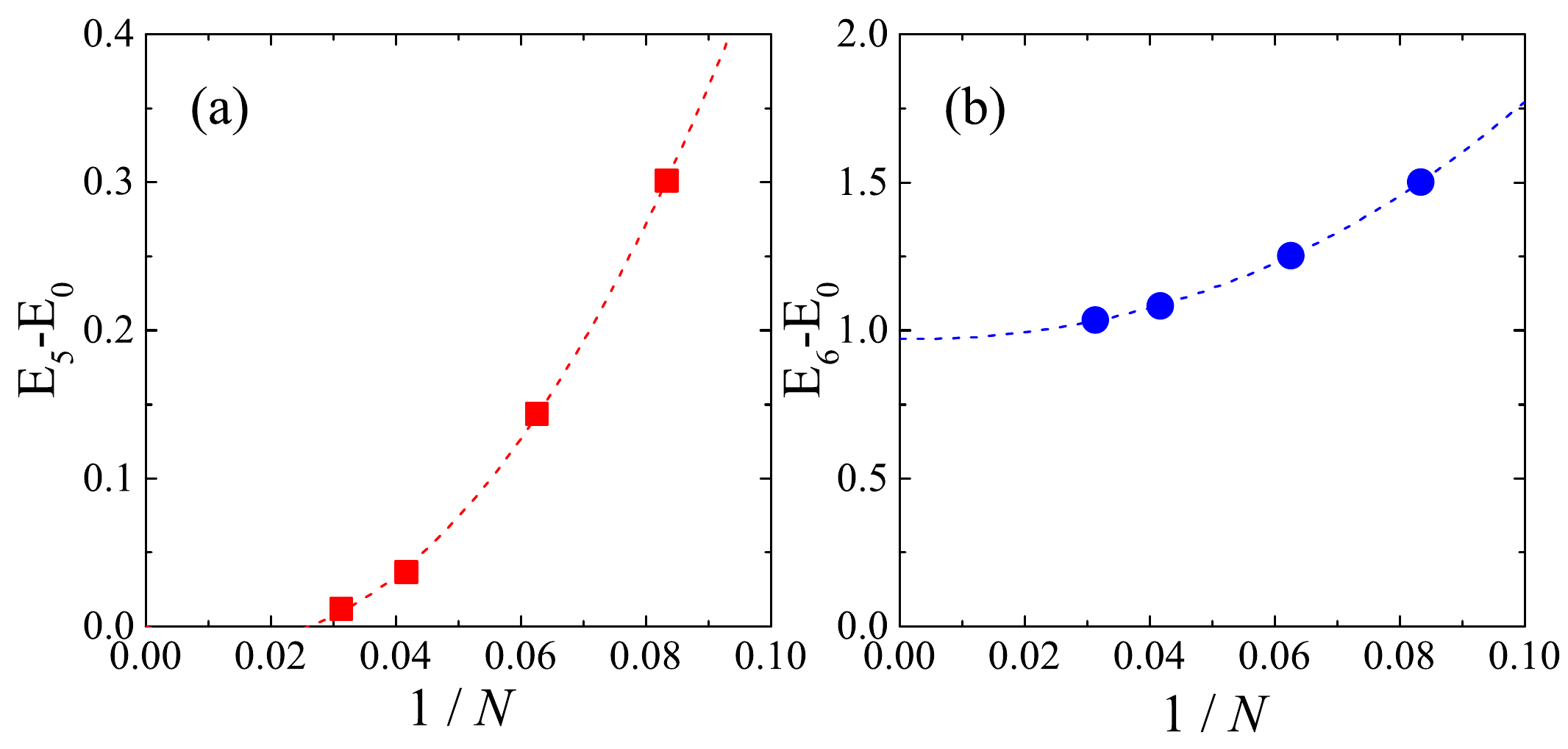}
  \caption{(a) The third-order polynomial extrapolations of (a) the finite-size interval of the six ground-state manifolds (GSM) and (b) the excitation gap above the GSM at $J_1^{\pm\pm}=0.6, J_1^{z\pm}=0, J_2=0$ [see Fig.~\ref{fig:AppStripeEngy} (a)]. }
  \label{fig:AppFitGap}
\end{figure}

We have calculated the low-energy spectra of different phases and find that there are six degenerate ground states in Stripe-I and Stripe-II phases, as shown in Fig.~\ref{fig:AppStripeEngy}. These six degenerate ground states are in the translation invariant momentum sectors $\Gamma, M_1, M_2, M_3$. Three of them are in the $\Gamma$ sector, while the other three distribute into three $M$ sectors. We can use finite-size scaling of energy gaps to verify the degeneracy in the thermodynamic limit which is shown in Fig.~\ref{fig:AppFitGap}.

Previous classical Monte Carlo study from Ref.~\onlinecite{LBalents2018} has shown that there are six basic spin-orbital-lock stripe configurations which differentiate by three choices of the principal lattice directions that stripes run along and two spin orientations within each stripe. For $J_1^{z\pm}=0$, in the stripe-I phase, the spins lay in the $xy$ plane and point perpendicular to the stripes (see Fig.~\ref{fig:PhaseDiagramI}), while in the stripe-II phase, the spins also lay in the $xy$ plane but point along the principal axes $\pm \mathbf{a}_1, \pm \mathbf{a}_2, \mp \mathbf{a}_1\pm \mathbf{a}_2$ (see Fig.~\ref{fig:PhaseDiagramI}). The nonzero $J_1^{z\pm}$ will tilt the spins out of $xy$ plane by an angle with the $z$ axis.

\section{Frustration parameter}
\label{App:Frustration}

The frustration parameter is defined as $f=|\Theta_{\rm{CW}}|/T_c$, where $\Theta_{\rm{CW}}$ is the negative Curie-Weiss temperature and $T_c$ is the critical temperature. We take the $T_c$ approximately as the temperature $T_{m}$ where the magnetic heat capacity gets its maximum value. Actually, $T_c\approx T_{m}$ works well in the stripe-I and stripe-II phases. However, in quantum spin liquid phase region, $T_c$ is zero. In fact, the frustration parameter should be diverge. And the heat capacity still has a broad maximum at finite $T$. In the 120$^{\circ}$ N\'eel phase, the $J_1^{\pm\pm}$ and $J_1^{z\pm}$ interactions break the $U(1)$ continuous symmetry of the XXZ model. Especially, the $J_1^{z\pm}$ interaction would tilt the spins out of $xy$ plane. Then whether the 120$^{\circ}$ N\'eel phase has a gap and a finite critical temperature are still unclear, which need further study in the future. In any case, we can expect that $T_c$ should be less than the $T_{m}$. Therefore, the frustration parameter in the 120$^{\circ}$ N\'eel phase is underestimate. Even though, using $T_c\approx T_{m}$ may not correctly estimate the actual frustration parameter. We still can use this approximation to compare the frustration of different phase regions in the 3D parameter space. As we have shown in the Fig.~\ref{fig:PhaseDiagramII} of the main text, the nonmagnetic quantum spin liquid region has a larger frustration parameter compared to other magnetic ordered phase regions, that is consistent with phase boundaries obtained by extrapolations of magnetic orders.

\begin{figure}[t]
  \centering
  \includegraphics[width=0.48\textwidth]{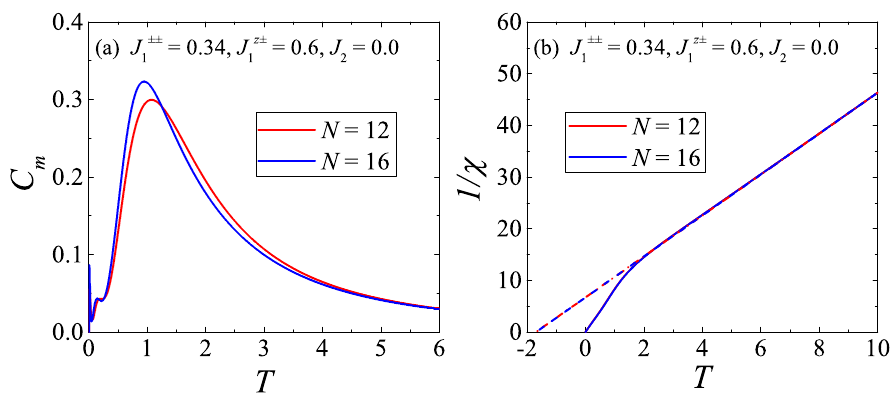}
  \caption{(a) Magnetic heat capacity and (b) uniform magnetic susceptibility obtained by full exact diagonalization using $12$ and $16$ clusters. For $12$ cluster, $T_m\approx 1.08, \Theta_{\rm{CW}}\approx -1.70, f\approx 1.57$. For $16$ cluster, $T_m\approx 0.95, \Theta_{\rm{CW}}\approx -1.70, f\approx 1.79$. We take the Boltzmann constant $k_B=1$ in drawing these two figures and use the B set of parameters to perform the calculations. Two prominent peaks are displayed in the heat capacity. The first peak in the low temperature comes from the finite-size gap of ground-state manifold (GSM). This peak will shift to zero temperature when the system size goes to infinite. The second peak reflects the finite excitation gap above the GSM. This peak will diverge when the system size goes to infinity, which indicates a spontaneously $Z_6$ symmetry breaking.}
  \label{fig:AppFrusParam}
\end{figure}

Here, we take the B set of parameters [see Fig.~\ref{fig:PhaseDiagramII}(a) in the main text] to representatively show the calculation of frustration parameter. The origin data of heat capacity and uniform magnetic susceptibility are shown in Fig.~\ref{fig:AppFrusParam}. These two observations are calculated by the following equations.
\begin{eqnarray*}
C_m &=&\frac{1}{Nk_B T^2}\left(\braket{H^2}-\braket{H}^2\right),\\
\chi &=&\frac{1}{Nk_B T}\left(\braket{M_z^2}-\braket{M_z}^2\right).
\label{Eq:Hmlt}
\end{eqnarray*}

\section{Magnetization curves}
\label{App:Magnetization}

\begin{figure}[t]
  \centering
  \includegraphics[width=0.478\textwidth]{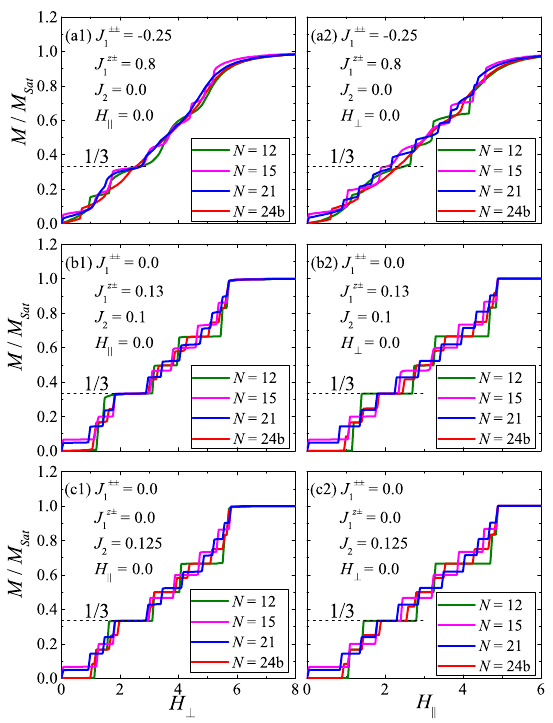}
  \caption{Magnetization curves of the QSL phase at different sets of parameters under the external magnetic fields. [(a1)--(c1)] The magnetic fields are perpendicular to the $c$ axis. [(a2)--(c2)] The magnetic fields are parallel to the $c$ axis. The set of parameters used in (b1) and (b2) corresponds to $C$ point in Fig.~\ref{fig:PhaseDiagramII}(c) of the main text.}
  \label{fig:AppMagFieldII}
\end{figure}

In this sector, we want to show more magnetization curves at different phases, including 120$^{\circ}$ N\'eel phase, Stripe-I phase, and quantum spin liquid phase. The magnetization curves with some sets of parameters in the quantum spin liquid region are representatively shown in Fig.~\ref{fig:AppMagFieldII}. In Figs.~\ref{fig:AppMagFieldII}(a1) and \ref{fig:AppMagFieldII}(a2), since the out-of-plane interaction $J_1^{z\pm}=0.8$ is large, it seems that the 1/3-magnetization plateau is already melted to be invisible, especially for the curve obtained by $24b$ cluster. And a more linear curve (in the thermodynamic limit) is observed when applying the field parallel to the $c$ axis. While for Figs.~\ref{fig:AppMagFieldII}(b1) and \ref{fig:AppMagFieldII}(c1), the $J_1^{z\pm}$ interaction is small or zero, so we can reproduce flat 1/3-magnetization plateaux.

\begin{figure}[t]
  \centering
  \includegraphics[width=0.478\textwidth]{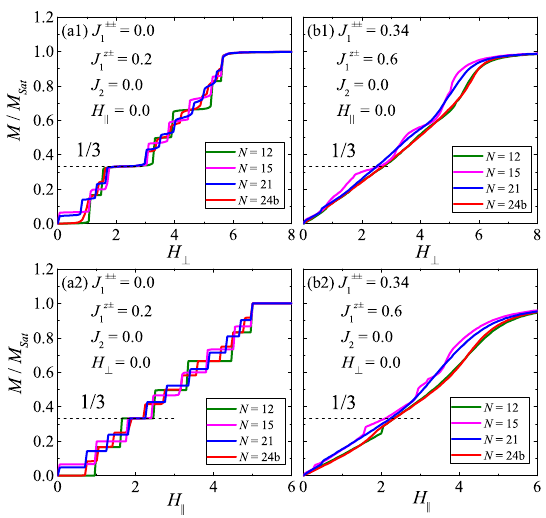}
  \caption{Magnetization curves at different sets of parameters under external magnetic fields. [(a1)--(b1)] The magnetic fields are perpendicular to the $c$ axis. [(a2)--(b2)] The magnetic fields are parallel to the $c$ axis. (a1) and (a2) are for the 120$^{\circ}$ N\'eel phase. (b1) and (b2) are for the stripe-I phase.}
  \label{fig:AppMagFieldIII}
\end{figure}

\begin{figure}[t]
  \centering
  \includegraphics[width=0.478\textwidth]{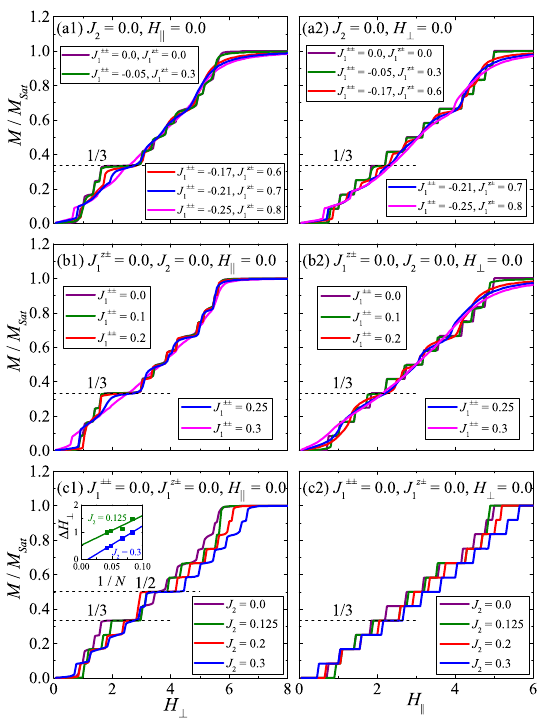}
  \caption{Finite-size magnetization curves obtained by $24b$ cluster along different paths in the 3D parameter space. [(a1)--(c1)] The magnetic fields are perpendicular to the $c$ axis. [(a2)--(c2)] The magnetic fields are parallel to the $c$ axis. The inset of (c1) shows linear extrapolations of the 1/3-plateau-width as functions of $1/N$.}
  \label{fig:AppMagFieldIV}
\end{figure}

\begin{figure}[htp!]
  \centering
  \includegraphics[width=0.478\textwidth]{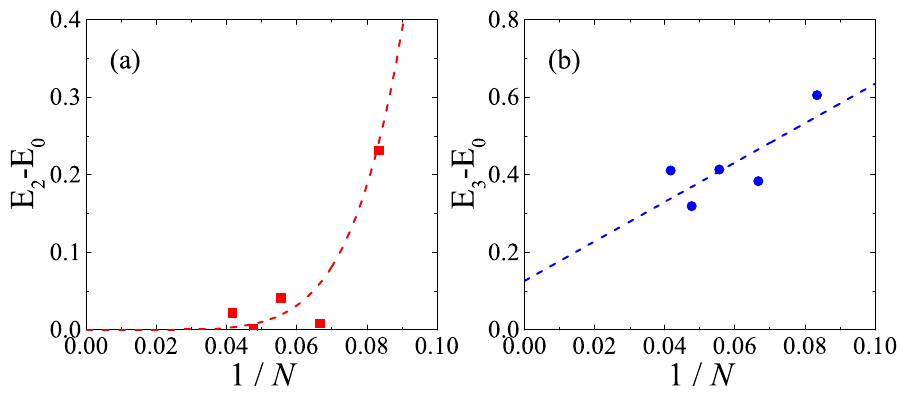}
  \caption{(a) Finite-size gap of the ground-state manifold as a function of $1/N$. The red dashed line is a guide to the eye. This gap will be zero when $N\rightarrow\infty$. (b) Linear extrapolation of the excitation gap above the ground-state manifold. The finite extrapolated value indicates a finite excitation gap. We take $J_1^{\pm\pm}=-0.17, J_1^{z\pm}=0.6, J_2=0.0, H_{\perp}=1.8\sqrt{2}, H_{\parallel}=0$ which corresponds to 1/3-magnetization plateau phase or $uud$ phase region in the ED calculation [see Fig.~\ref{fig:MagFieldI}(a) of the main text].}
  \label{fig:AppuudGap}
\end{figure}

We have also calculated the magnetization curves of 120$^{\circ}$ N\'eel phase and Stripe-I phase in Fig.~\ref{fig:AppMagFieldIII}. In the 120$^{\circ}$ N\'eel phase, the 1/3-magnetization plateau is clearly seen. The nonflatness depends on the $J_1^{z\pm}$ interaction. In the Stripe-I phase, there is no 1/3-magnetization plateau induced by two kinds of magnetic fields.

\begin{figure}[t]
  \centering
  \includegraphics[width=0.5\textwidth]{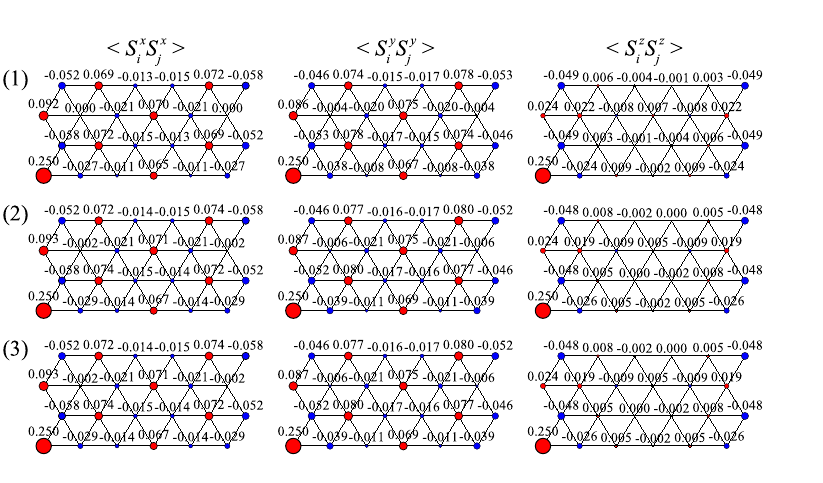}
  \caption{Three components of the spin correlation functions at $J_1^{\pm\pm}=-0.17, J_1^{z\pm}=0.6, J_2=0.0, H_{\perp}=1.8\sqrt{2}, H_{\parallel}=0$ [see Fig.~\ref{fig:MagFieldI}(a) in the main text].}
  \label{fig:AppPCorr}
\end{figure}

To show the effects of different exchange interactions, like $J_1^{\pm\pm},J_1^{z\pm},J_2$, on the 1/3-magnetization plateau, we use $24b$ cluster to show the change of magnetic curves with these parameters, which are shown in Fig.~\ref{fig:AppMagFieldIV}. When $J_1^{\pm\pm},J_1^{z\pm}, J_2$ are small and the system is in 120$^{\circ}$ N\'eel phase, the 1/3-magnetization plateau is flat. In the quantum spin liquid phase region with large $J_1^{z\pm}>0.5$, the 1/3-magnetization plateau is melted to nonlinear rough curve, see Fig.~\ref{fig:AppMagFieldIV}(a1). When we increase $J_1^{\pm\pm}$ and keep $J_1^{z\pm}=0$, the flatness of plateau is nearly unchanged. After $J_1^{\pm\pm}>0.2$ which drives system into Stripe-I phase, the plateau quickly melts to a linear curve, see Fig.~\ref{fig:AppMagFieldIV}(b1). For the $J_1-J_2$ XXZ model, in the 120$^{\circ}$ N\'eel and the QSL phase regions, the 1/3-magnetization plateau is flat and has nonzero width $\Delta H_{\perp}$ in the thermodynamic limit [see the inset of Fig.~\ref{fig:AppMagFieldIV}(c1)]. When $J_2>0.175$, the system is in the stripe phase with threefold ground-state degeneracy, the 1/3-magnetization plateau disappears [see the inset of Fig.~\ref{fig:AppMagFieldIV}(c1)]. Instead, a 1/2-magnetization plateau appears. 

To verify the 1/3-magnetization plateau phase is a $uud$ phase. We have calculated the energy spectrum and the spin correlation functions at $J_1^{\pm\pm}=-0.17, J_1^{z\pm}=0.6, J_2=0.0, H_{\perp}=1.8\sqrt{2}, H_{\parallel}=0$, see Fig.~\ref{fig:AppuudGap}. From the low-energy spectrum, we find threefold quasidegenerate ground states. Through finite-size scalings, we can observe the exact degeneracy (before spontaneously $Z_3$ symmetry breaking) and a finite-energy gap above the ground-state manifold. And we show the real-space spin correlation functions of these three ground states in Fig.~\ref{fig:AppPCorr}, the $uud$ structure can be clearly seen.

\section{VII: Bond randomness effects}

\begin{figure}[t]
  \centering
  \includegraphics[width=0.44\textwidth]{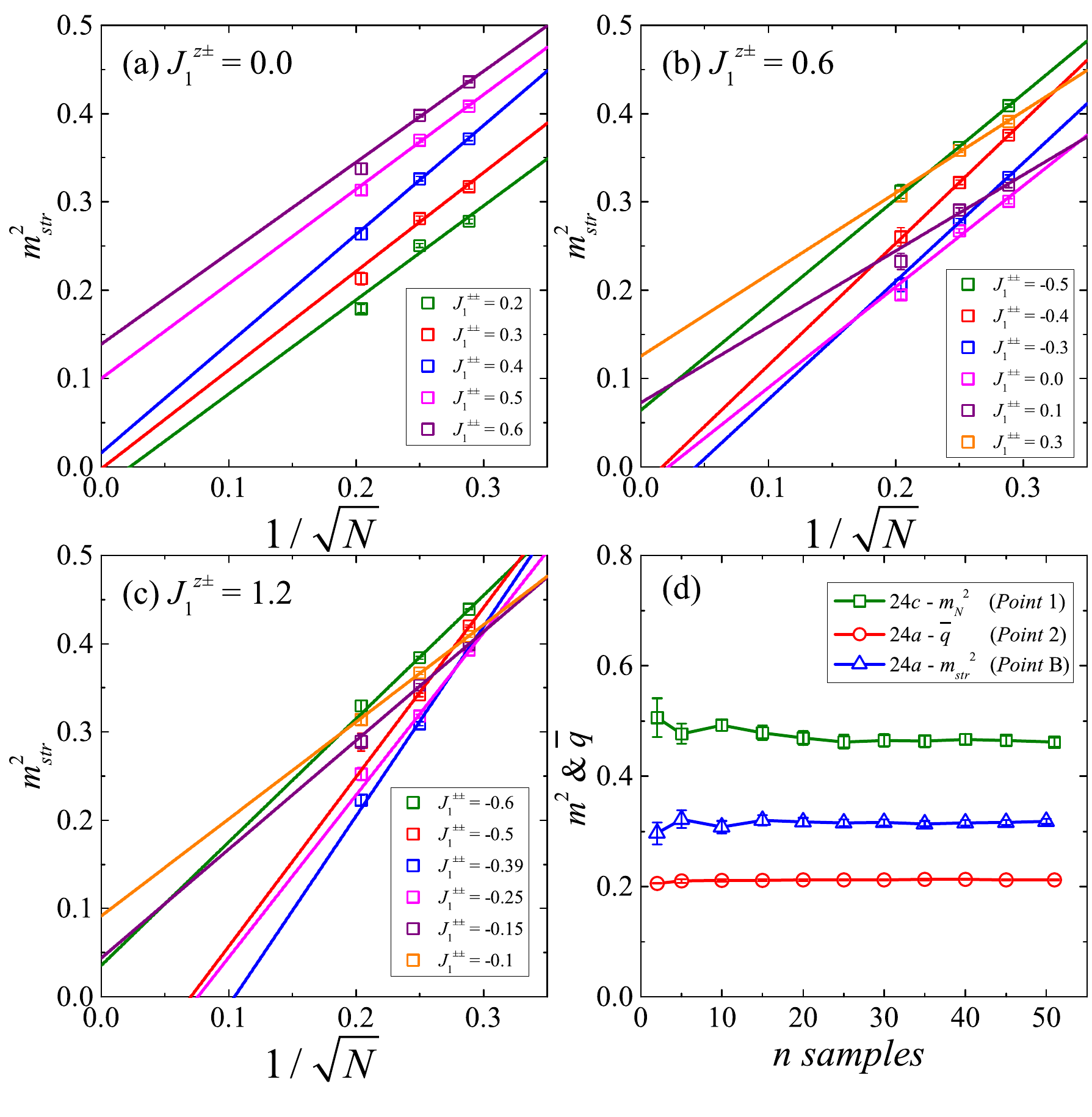}
  \caption{The linear extrapolations of the square sublattice magnetization for the stripe order parameters in selective paths which go along (a) $J_1^{z\pm}=0.0$, (b) $J_1^{z\pm}=0.6$, and (c) $J_1^{z\pm}=1.2$ horizontal lines in Fig.~\ref{fig:GlassOrder}(a) of the main text. Panel (d) shows the change of three order parameters with the increasing of random samples.}
  \label{fig:AppRMagOrdII}
\end{figure}

To simulate chemical disorders in YbMgGaO$_4$, we have introduced bond randomness into the Hamiltonian. And there are four sets of parameters have been frequently used to do the calculations, $J_1^{\pm\pm}=0.0, J_1^{z\pm}=0.2, J_2=0.0~(Point~1)$; $J_1^{\pm\pm}=-0.2, J_1^{z\pm}=0.7, J_2=0.0~(Point~2)$; $J_1^{\pm\pm}=-0.35, J_1^{z\pm}=0.6, J_2=0.0~(Point~3)$; and $J_1^{\pm\pm}=0.341, J_1^{z\pm}=0.598, J_2=0.0~(Point~B)$. These four sets of parameters have been marked in Fig.~\ref{fig:GlassOrder}(a) of the main text. For the $120^{0}$ N\'eel phase, the strongest randomness at $\Delta=1$ can eliminate this magnetic order, which can be seen in Fig.~\ref{fig:GlassOrder}(b) of the main text. For most of the stripe-phase region, the stripe orders are stable against the bond randomness and cannot be eliminated even in the strongest bond-randomness case, as can be seen in Fig.~\ref{fig:GlassOrder}(b) of the main text. And for QSL phases, both in the clean case and the strongest bond randomness limit, the vanishing of the average spin freezing order parameter [shown in Fig.~\ref{fig:MagOrders}(d) and Fig.~\ref{fig:GlassOrder}(b) of the main text] indicates the absence of spin-glass order. To confirm the convergence, we show different order parameters changing with the number of random samples in Fig.~\ref{fig:AppRMagOrdII}(d). We are confident that using at least 20 bond-randomness samples is able to get reliable randomly averaged order parameters.

\begin{figure*}[htp!]
  \centering
  \includegraphics[width=0.925\textwidth]{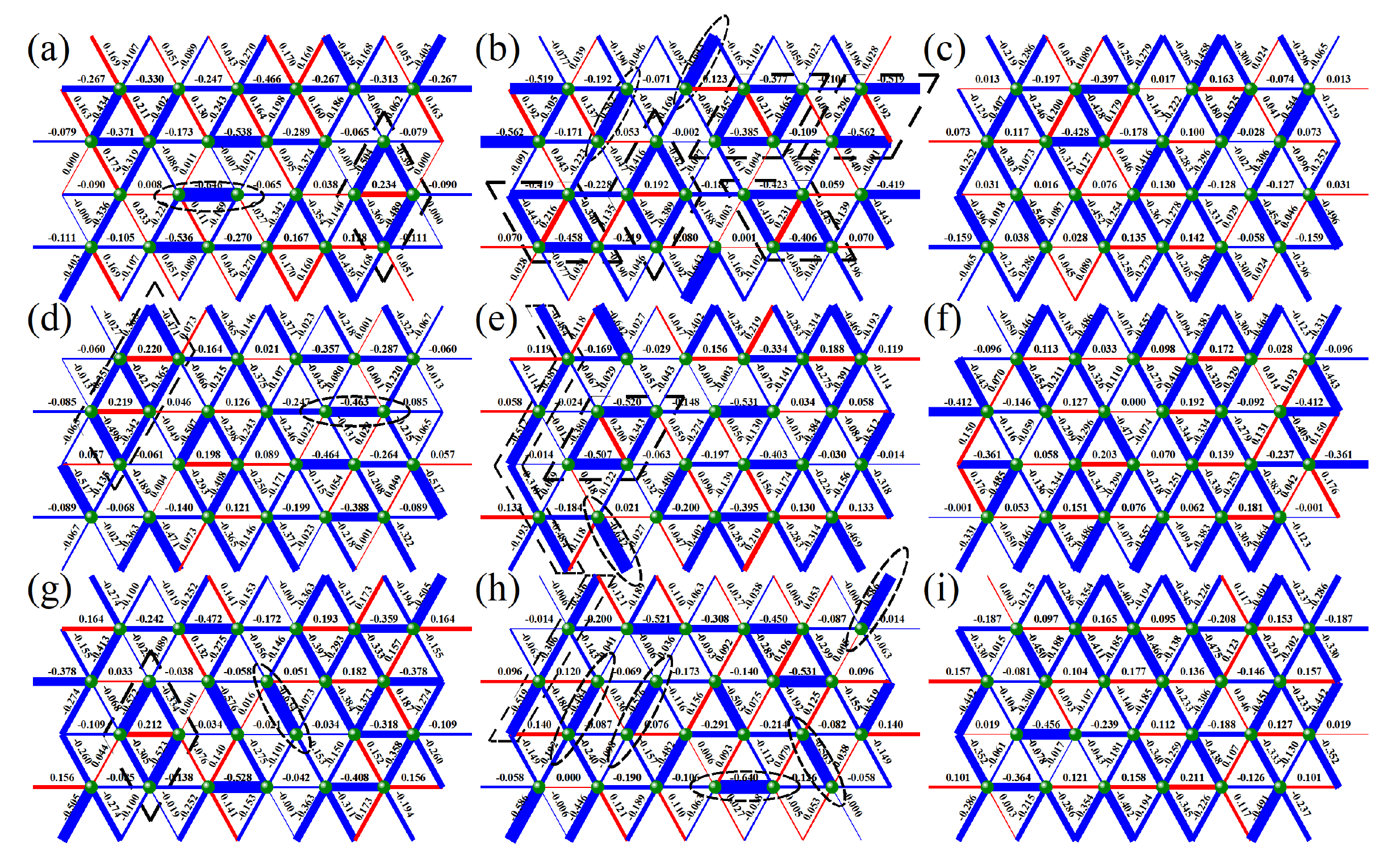}
  \caption{The nearest-neighbor spin correlations for different random configurations under the strongest bond-randomness $\Delta=1.0$ at $Point$ 2 [see Fig.~\ref{fig:GlassOrder}(a) in the main text]. Some distinct two-spin singlets and other singlet clusters are marked by dotted box.}
  \label{fig:AppQSLRNNCorr}
\end{figure*}

Using linear extrapolations of the stripe order parameter shown in Figs.~\ref{fig:AppRMagOrdII}(a)--\ref{fig:AppRMagOrdII}(c), we obtain the phase diagram under the strongest bond randomness $\Delta$ = 1.0, which is shown in Fig.~\ref{fig:GlassOrder}(a) of the main text. In the nonmagnetic spin-liquid-like (SLL) phase region, we also show the average spin freezing order parameter to rule out the spin glass phase. This SLL phase is actually a 2D analog of random-singlet (RS) phase. To see more clear about this phase, we plot the real-space spin correlations under some representative bond randomness configurations, which are shown in Fig.~\ref{fig:BNNCorr} of the main text and Fig.~\ref{fig:AppQSLRNNCorr}.

\begin{figure}[t]
  \centering
  \includegraphics[width=0.46\textwidth]{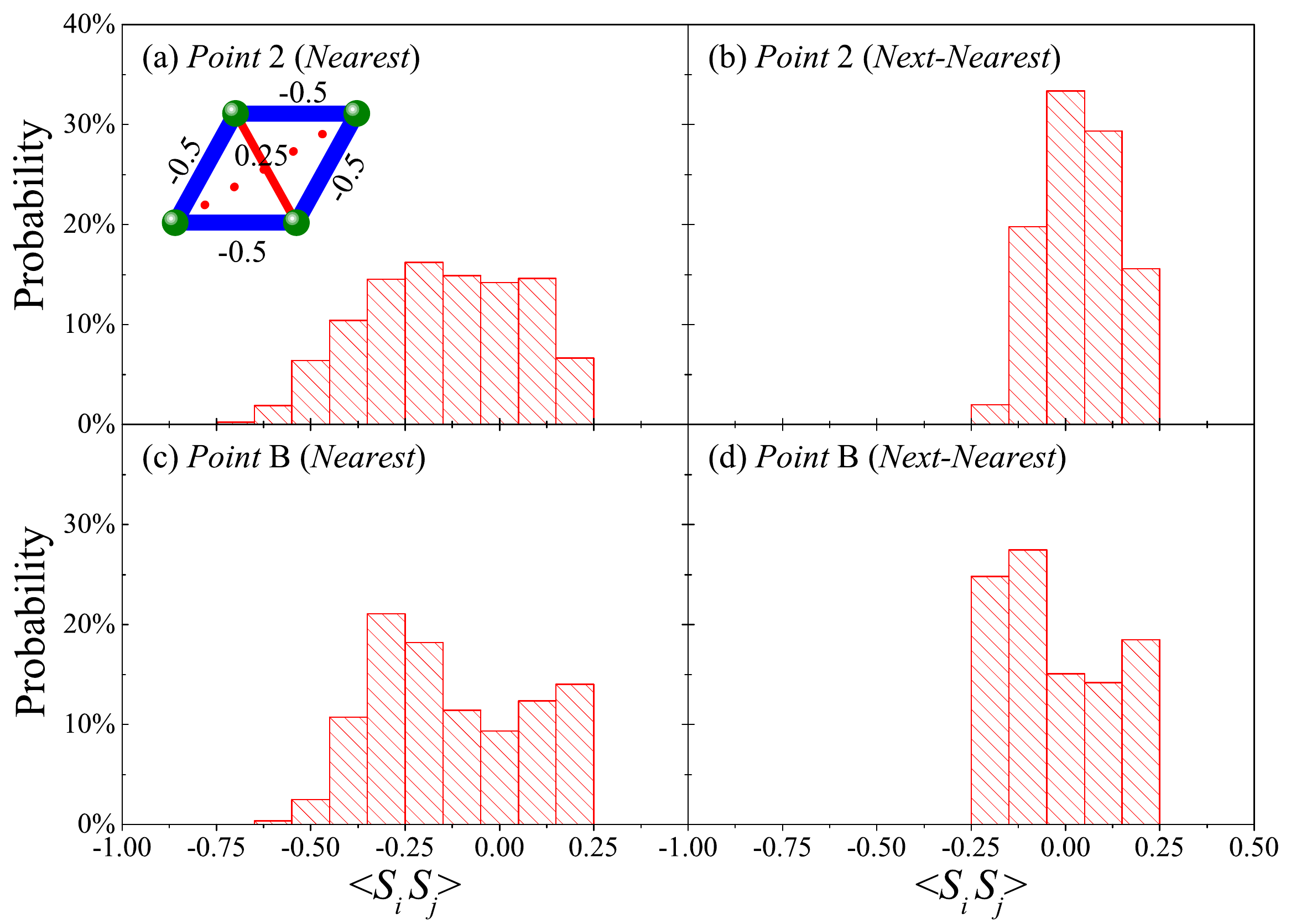}
  \caption{Histograms of nearest-neighbor and next-nearest-neighbor spin correlations at [(a), (b)] $Point$ 2 and [(c), (d)] $Point$ B under the strongest bond randomness limit $\Delta$ = 1.0. We use $24b$ cluster and 153 bond-randomness samples to get those histogram. The inset of (a) is the nearest-neighbor and next-nearest-neighbor spin correlations of a four-spin singlet. If four-spins form an exact singlet (plaquette singlet), the nearest-neighbor spin correlations with blue lines would be $-0.5$, while spin correlations with red solid and dashed lines are $0.25$.}
  \label{fig:AppQSLRHistogram}
\end{figure}

In the RS phase, we can find some random distributions of nearest-neighbor two-spin singlets, four-spin singlets and other larger singlet domains. If two nearest-neighbor spins form an exact singlet, then the spin correlation between these two spins is equal to $-0.75$. However, due to the geometry frustration and competition between nearest-neighbor bonds sharing one of the same lattice site, two nearest-neighbor spins can approximately form a local singlet if their correlation is close to $-0.75$. In Fig.~\ref{fig:BNNCorr} of the main text and Fig.~\ref{fig:AppQSLRNNCorr}, we representatively show the two-spin singlets (or dimers) which are marked by the dotted oval boxes. For four nearest-neighbor spins forming a (plaquette) singlet, the spin correlations between diagonal sites [red solid and red dashed lines in Fig.~\ref{fig:AppQSLRHistogram}(a)] are equal to $0.25$ which represents ferromagnetic correlation, that will contribute to the nonzero fraction of ferromagnetic correlations in the histogram of Figs.~\ref{fig:AppQSLRHistogram}(a) and \ref{fig:AppQSLRHistogram}(b). Similarly, we have also found six-spin singlets which are representatively shown in Fig.~\ref{fig:BNNCorr}(a2) of the main text and Fig.~\ref{fig:AppQSLRNNCorr}(d). Other larger singlet domains can also be found. But how do we define a spin domain? For two spins with their correlation larger than $-0.25$, they are disentangled in the $SU(2)$ limit~\cite{Kawamura2019}. Therefore, we can set the criterion of two spins belonging to the same domain as the spin correlation between them is less than $-0.25$ (see more details in Ref.~\onlinecite{Kawamura2019}). In Fig.~\ref{fig:GlassOrder}(d) of the main text, we show the histogram of different spin domains in the RS phase. The dominant contributions are the two-spin singlets and other larger singlet domains with even number of spins. In Figs.~\ref{fig:AppQSLRHistogram}(a) and \ref{fig:AppQSLRHistogram}(b), we show the distribution of nearest-neighbor spin correlations to see more details from another aspect. As we known, if two nearest-neighbor spins form a nearly singlet, then the spin correlations of other ten nearest-neighbor bonds sharing one of these two spins in the triangular lattice will be very weak. Therefore, unlike the 1D bipartite Heisenberg chain, due to the large coordination number and the geometry frustration of triangular lattice, the percentage of singlet bonds in all nearest-neighbor bonds will be small, as can be seen in Fig.~\ref{fig:AppQSLRHistogram}(a). In the formation of four-spin singlet, the spin correlations of diagonal nearest-neighbor and next-nearest-neighbor bonds (for triangular lattice) are $0.25$ which contributes to the nonzero fraction of ferromagnetic correlations in Figs.~\ref{fig:AppQSLRHistogram}(a) and \ref{fig:AppQSLRHistogram}(b).

In the Stripe-I phase, take B set of parameters ($Point~B$) as example, we show the histogram of nearest-neighbor and next-nearest-neighbor spin correlations in Figs.~\ref{fig:AppQSLRHistogram}(c) and \ref{fig:AppQSLRHistogram}(d), respectively. The bond randomness cannot fully destroy the stripe order. Therefore, we can see a large fraction of antiferromagnetic correlation ($\braket{\mathbf{S}_i\mathbf{S}_j}\sim$ $-$0.25) and ferromagnetic correlation ($\braket{\mathbf{S}_i\mathbf{S}_j}\sim$ 0.25) in the nearest-neighbor and the next-nearest-neighbor spin correlations.

\begin{figure}[b]
  \centering
  \includegraphics[width=0.455\textwidth]{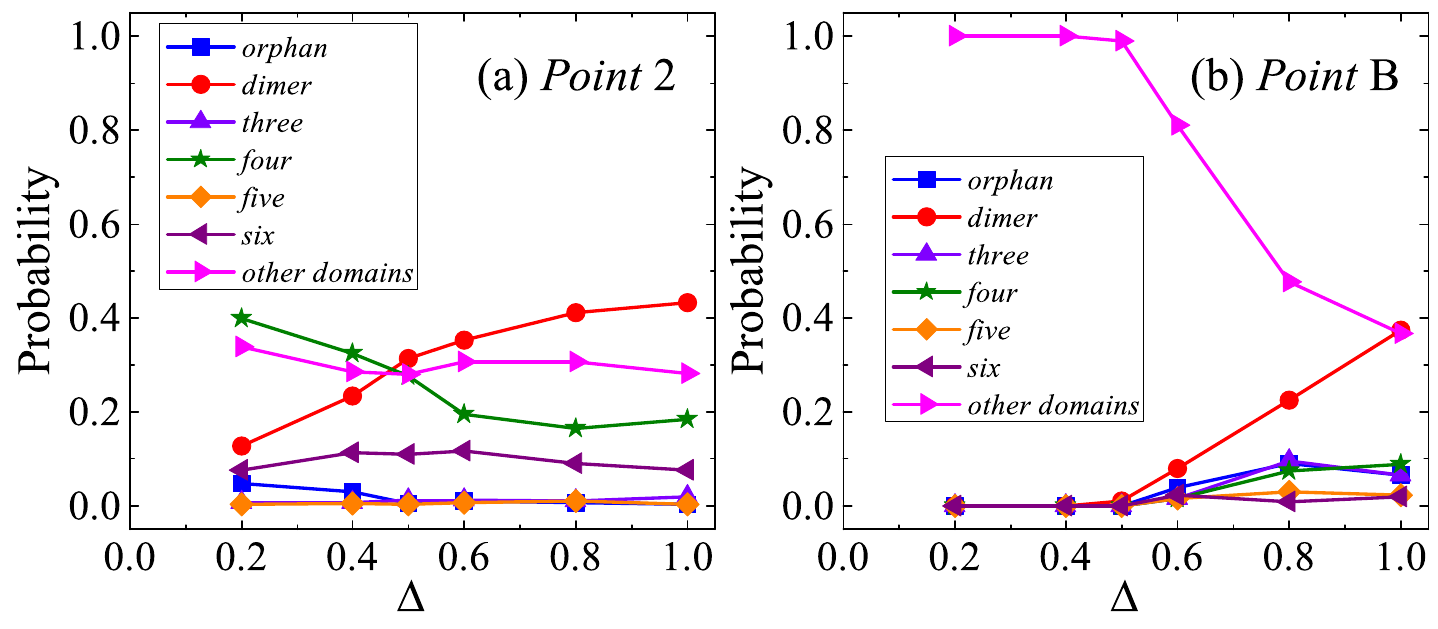}
  \caption{Distributions of spin domains as functions of $\Delta$ at (a) $Point$ 2 and (b) $Point$ B. We use 24b cluster and at least 100 bond-randomness samples to get these distributions.}
  \label{fig:AppQSLRHistogramII}
\end{figure}

To show how the bond randomness strength affects the ground state, we show the distribution of spin domains as a function of $\Delta$ in Fig.~\ref{fig:AppQSLRHistogramII}. At $Point$ B which is shown in Fig.~\ref{fig:AppQSLRHistogramII}(b), in the weak bond randomness regime with $\Delta \leq$  0.5, because of the large excitation gap, nearly all the spins are in one domain with stripe ordering. With the increasing of the randomness, say $\Delta \geq$ 0.5, large domains are gradually broken into some smaller domains, especially like the two-spin singlets or dimers. While at $Point$ 2 without any magnetic ordering in the clean case, spins can easily form some small singlet domains even under weak bond randomness ($\Delta$ = 0.2) [shown in Fig.~\ref{fig:AppQSLRHistogramII}(a)]. Interestingly, spins prefer to form four-spin singlets or resonating dimers when $\Delta < 0.5$ instead of two-spin singlets or dimers which dominate the case of stronger randomness ($\Delta \geq$ 0.5). For the orphan or isolated spin, its fraction is nonnegligible when $\Delta<0.5$ (4.8\% at $\Delta$ = 0.2 and 3.0\% at $\Delta$ = 0.4) and then drops to below 1\% when $\Delta \geq$ 0.5.

\begin{figure}[t]
  \centering
  \includegraphics[width=0.455\textwidth]{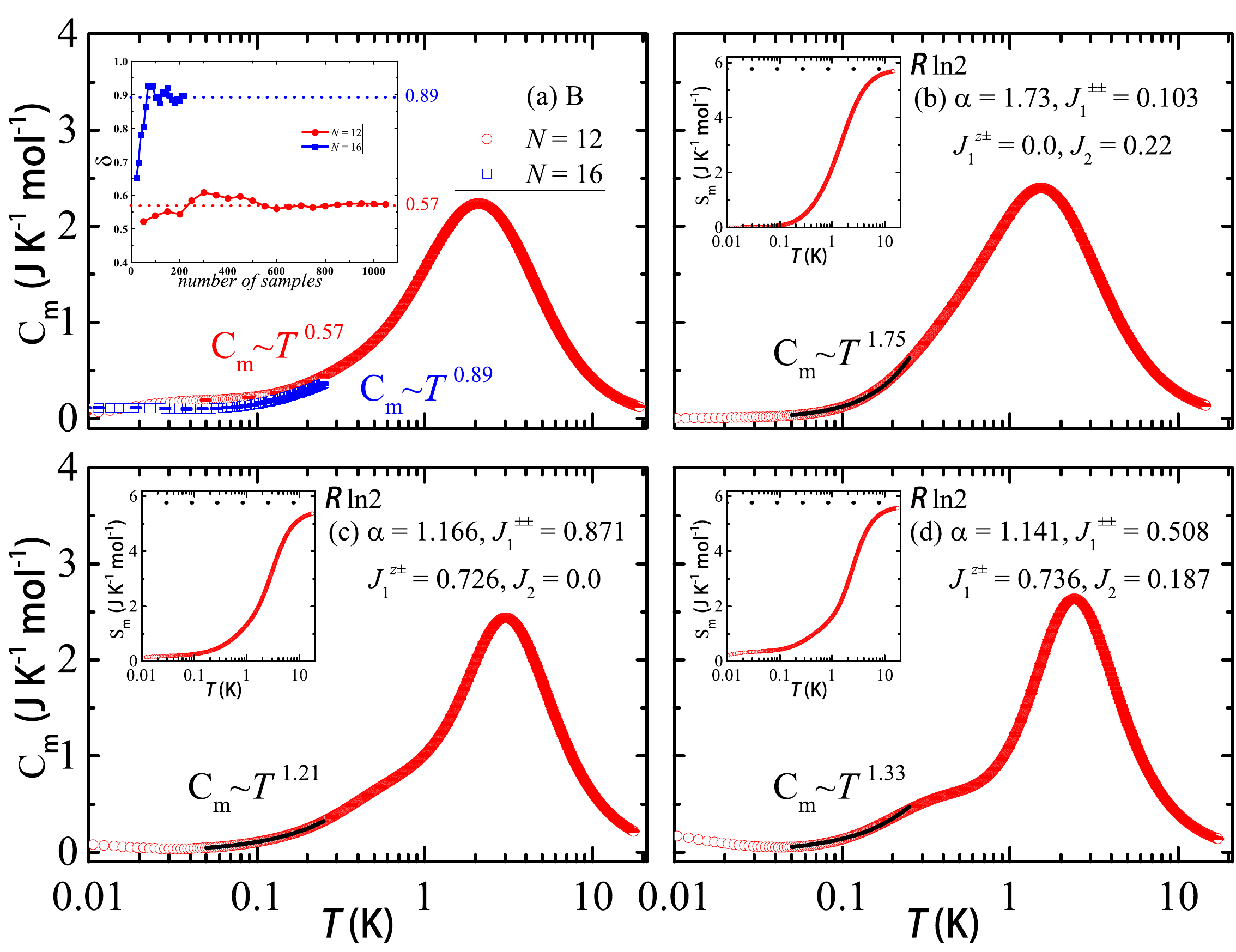}
  \caption{(a) The magnetic heat capacities $C_m$ obtained by $12$ and $16$ clusters in the strongest bond randomness limit $\Delta=1$ for the B set of parameters: i.e., $J_1^{\pm\pm}/J_1=0.34, J_1^{z\pm}/J_1=0.6, J_2/J_1=0$, and we use $J_1=0.164$ meV~\cite{YDLi2018} to do the ED calculations. For the $16$ cluster, we employ Lanczos method to calculate the heat capacity at low temperature. The restriction of Boltzmann factor $e^{-(E_{\rm{max}}-E_0)/k_B T}<10^{-12}$ has been used to determine the upper-bound temperature below which the calculated $C_m$ is trustable. The inset shows the change of exponent $\delta$ obtained by power-law fitting the $C_m$ curve from $T = 0.05$ K to $T = 0.25$ K with the increasing of the number of samples. [(b)--(d)] Magnetic heat capacities $C_m$ obtained by $12$ cluster in the strongest bond randomness limit $\Delta=1$ for other three sets of parameters fitting by the experiments. We use $J_1=$ (b) 0.126 meV~\cite{Paddison2017}, (c) 0.1515 meV, and (d) 0.1495 meV~\cite{XinshuZang2018} to show all the data. We have used at least 600 bond-randomness samples for $12$ cluster and at least 220 bond-randomness samples for $16$ cluster to get the averaged $C_m(T)$. The insets show the magnetic entropy $S_m=\int_0^{T} C_m/T dT$.}
  \label{fig:AppExperimentCv}
\end{figure}

\begin{figure}[t]
  \centering
  \includegraphics[width=0.455\textwidth]{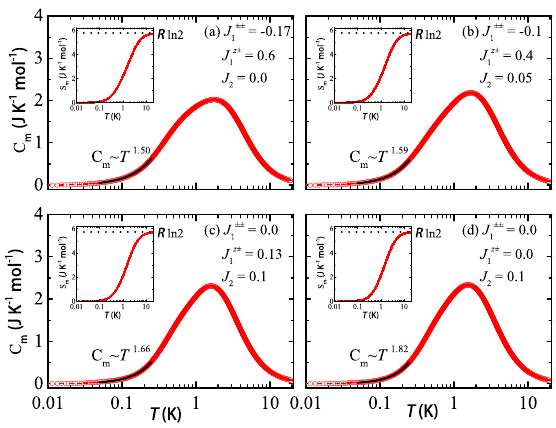}
  \caption{Magnetic heat capacities $C_m$ obtained by $12$ cluster in the strongest bond randomness limit $\Delta=1$. Here, we use $J_1=0.164$ meV to show all the data. And we have used at least 600 bond-randomness samples to get the averaged $C_m(T)$. The insets show the magnetic entropy $S_m=\int_0^{T} C_m/T dT$.}
  \label{fig:AppQSLCv}
\end{figure}
\begin{figure}[t]
  \centering
  \includegraphics[width=0.45\textwidth]{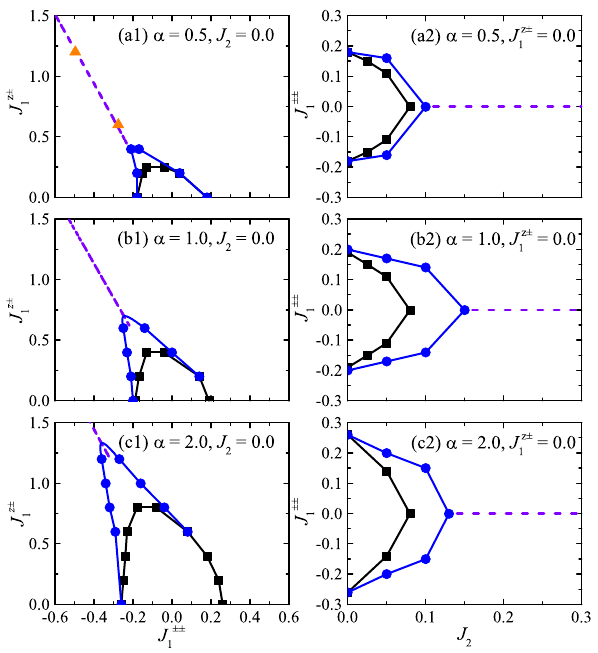}
  \caption{The phase diagrams with different easy-plane anisotropic $\alpha$. [(a1)--(c1)] are the phase diagrams on the $J_2=0, J_1^{\pm\pm}-J_1^{z\pm}$ plane. [(a2)--(c2)] are the phase diagrams on the $J_1^{z\pm}=0, J_2-J_1^{\pm\pm}$ slice. The blue phase transition points are obtained from linear extrapolations of the stripe order parameters, while the black ones are obtained from the peak position of fidelity susceptibility using $24a$ cluster. The purple dashed lines are the classical phase transition lines between two stripe phases, and the yellow points in (a1) are obtained by the level crossings of low excited energy states using $24a$ cluster.}
  \label{fig:AppAlpha}
\end{figure}

The above discussions focus on ground-state properties at zero temperature. Here, we want to discuss the bond randomness effects at finite temperature. Figure~\ref{fig:AppExperimentCv}(a) shows the magnetic heat capacity $C_m$ of B set of parameters which have also been shown in Fig.~\ref{fig:GlassOrder}(c) of the main text. And we have used sufficient random samples to make the power-law exponent $\delta$ converged, which can be seen in the inset of Fig.~\ref{fig:AppExperimentCv} (a).

In the strong randomness case, the finite-size effect is actually not severe. So the $12$ cluster is able to capture the main physics in the strongest bond-randomness limit. In this limit, the heat capacity has a broad peak, and this peak will not diverge with the increasing system size. That means even though the ground state of the system has residual stripe order, but it may be hard to probe this order at finite temperature. Actually, previous classical Monte Carlo simulation from Ref.~\onlinecite{LBalents2018} has shown the similar behavior in the heat capacity. In the clean case, there is a single critical temperature with slowly diverging heat capacity. In the randomness case, this transition is removed by fragmenting the system into spin domains.

We have also calculated the heat capacity with other sets of parameters (especially for the sets of parameters fitting by different research groups or within the QSL phase region) under the strongest bond randomness. However, none of those can reproduce nearly the same heat capacity as the experimental one, which are show in Figs.~\ref{fig:AppExperimentCv} and \ref{fig:AppQSLCv}. In the clean limit, whether we can get the same heat capacity as the experimental one using some sets of parameters is still an open question. Recently, a finite-temperature Lanczos methods with improved accuracy has successfully applied to Kitaev-Heisenberg model on Kagome and triangular lattices~\cite{Morita2020}, which would be a great help to study the finite-temperature properties of the model related to YbMgGaO$_4$ in future.

\section{XXZ anisotropic effects}
\label{App:XXZ}

\begin{figure}[t]
  \centering
  \includegraphics[width=0.44\textwidth]{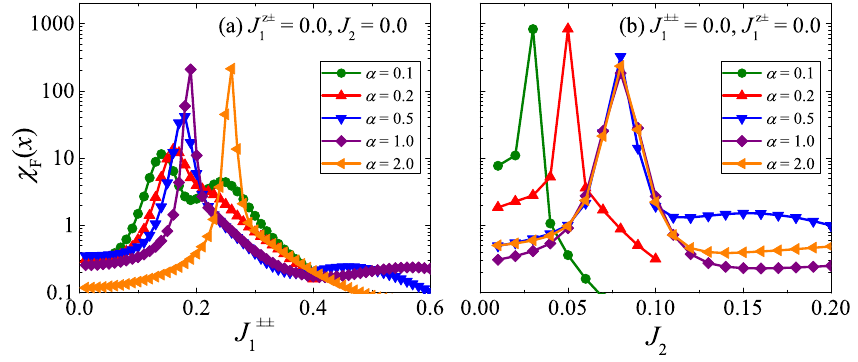}
  \caption{(a) The fidelity susceptibilities as functions of $J_1^{\pm\pm}$ under different XXZ anisotropic $\alpha$ and along the $J_1^{\pm\pm}$ axis. (b) The fidelity susceptibilities as functions of $J_2$ under different XXZ anisotropic $\alpha$ and along the $J_2$ axis. When $\alpha<0.5$, the phase transition point $J_{2,c}$ starts to drop quickly. Here, we use $24a$ cluster to perform the calculations.}
  \label{fig:AppAFidelity}
\end{figure}

To see how the XXZ (or easy-plane) anisotropic $\alpha$ affects the phase diagram, we use fidelity susceptibility of $24a$ cluster to get the 120$^{\circ}$ N\'eel phase boundaries under different $\alpha$, and use the linear extrapolations of the stripe order (mainly using $16$ and $24a$ clusters) to get the phase boundaries of stripe phases. Then we obtain some phase diagrams under different $\alpha$ which are shown in Fig.~\ref{fig:AppAlpha}. When $\alpha$ decreases, the (deformed) 120$^{\circ}$ N\'eel phase and the QSL phase regions shrink. Especially for the QSL phase, at $\alpha=0.5$, this phase region is too small to identify. Therefore, $\alpha_c\sim 0.5$ is a approximate critical value where the QSL disappears. In the limit of $\alpha=0$, the 120$^{\circ}$ N\'eel phase region will quickly vanish [see Fig.~\ref{fig:AppAFidelity} (b)]. When $\alpha$ is large, both of the 120$^{\circ}$ N\'eel phase and the QSL phase seem to extend to larger areas. Please remind that we have taken $J_1=1$ (actually $J_1^{zz}=1$) as the energy unit. If we take $J_1^{\pm}=1$ as the new energy unit, then the area of QSL phase region may decrease to a finite constant when we increase $\alpha$ from 1 to larger values. In the limit $\alpha=\infty$ or $J_1^{zz}=J_2^{zz}=0$, unlike the $\alpha=0$ limit, the quantum spin liquid phase will survive~\cite{suzuki2019}. Compared with previous DMRG study from Ref.~\onlinecite{ZYZhu2017} and Ref.~\onlinecite{ZYZhu2018}, our QSL regions are more naturally located between three magnetic ordered phases due to the order-by-disorder effect and extend to the $J_1^{\pm\pm}$ axis in the $J_2-J_1^{\pm\pm}, J_1^{z\pm}=0$ plane.

\section{Static and dynamical spin structure factor}

\begin{figure}[htp!]
  \centering
  \includegraphics[width=0.49\textwidth]{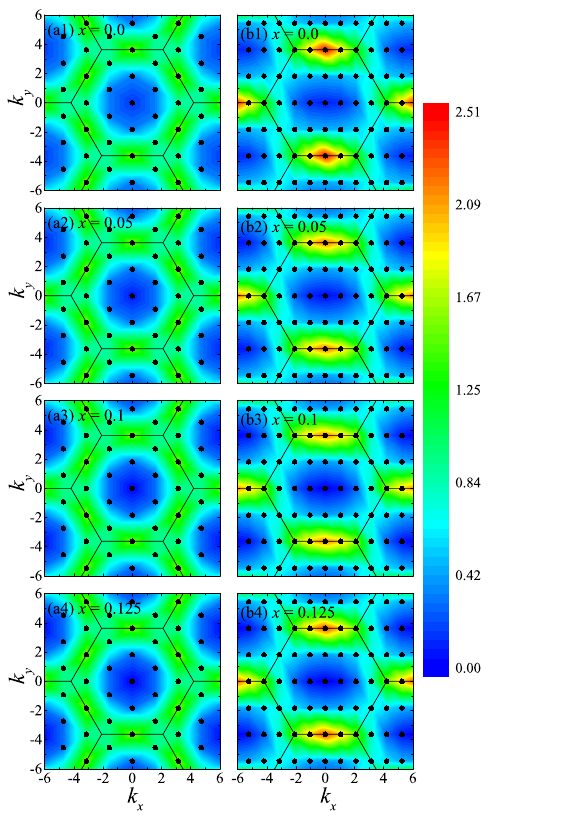}
  \caption{Spin structure factors at different $x$. $x$ specifies the exchange interactions $J_2=x, J_1^{\pm\pm}=1.6x-0.2, J_1^{z\pm}=-5.6x+0.7, x\in[0,0.125]$. [(a1)--(a4)] are obtained by $16$ cluster, while (b1--b4) are obtained by $24b$ cluster.}
  \label{fig:AppMagStrFct}
\end{figure}

The inelastic neutron scattering experiment of YbMgGaO$_4$ has revealed a broad low-energy excitation maxima at the M point and the concentrated spectral weight at the boundary of Brillouin zone. Here we show the contour plots of the static spin structure factors of the QSL region in Fig.~\ref{fig:AppMagStrFct}. We take a straight-line path $J_2=x, J_1^{\pm\pm}=1.6x-0.2, J_1^{z\pm}=-5.6x+0.7, x\in[0,0.125]$ in the 3D parameter space to show the static spin structure factors of QSL phase. The broad peaks at the M points signature short-range stripelike spin correlations in the QSL phase.

\begin{figure}[t]
  \centering
  \includegraphics[width=0.45\textwidth]{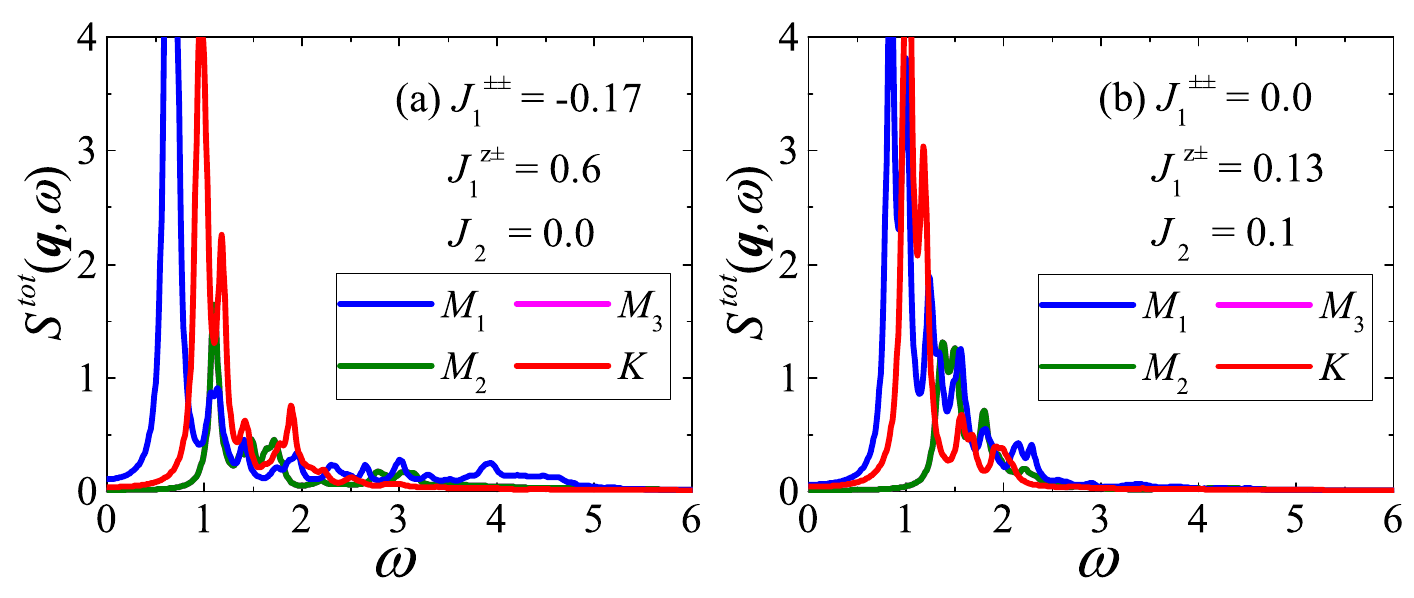}
  \caption{The dynamical spin structure factors of the QSL phase at two K points and three M points using different sets of parameters. The XXZ anisotropic $\alpha$ is set to be 1.317. Three M points are not equivalent in $24b$ cluster which we use here to do the calculations. The parameters we choose for panel (a) are $J_1^{\pm\pm}=-0.17, J_1^{z\pm}=0.6, J_2=0.0$. And panel (b) uses the C set of parameters which is shown in Fig.~\ref{fig:PhaseDiagramII}(c) of the main text. The Lorentz broadening factor we use is $\eta$ = 0.05.}
  \label{fig:AppQSLSqw}
\end{figure}

We also calculate the dynamical spin structure which can be studied by inelastic neutron scattering (INS) or x-ray Raman scattering in experiment. The dynamical spin structure factor in the QSL region can be computed by continued fraction expansion,
\begin{equation*}
\begin{split}
S^{tot}(\mathbf{q},\omega) &= S^{xx}(\mathbf{q},\omega) + S^{yy}(\mathbf{q},\omega) + S^{zz}(\mathbf{q},\omega),\\
S^{\alpha\alpha}(\mathbf{q},\omega)&=\sum_{n}\left\{|\langle \psi_n | \hat{S}^\alpha_{\bf q} | \psi_0 \rangle|^2 \delta\left[\omega-(E_n-E_0)\right]\right\}\\
  &=-\frac{1}{\pi}\lim\limits_{\eta\rightarrow 0}\text{Im}\left[\cfrac{\bra{\psi_{0}}\left(\hat{S}_{\mathbf{q}}^{\alpha}\right)^{\dagger}\hat{S}_{\mathbf{q}}^{\alpha}\ket{\psi_{0}}}
  {z-a_{0}-\cfrac{b_{1}^{2}}{z-a_{1}-\cfrac{b_{2}^{2}}{z-a_{2}\cdots}}}\right],
\end{split}
\end{equation*}
where $\alpha=x,y,z$ label the spin indices, $z=\omega+i\eta+E_0$, $a_{i}$ and $b_{i+1}$ are the diagonal and subdiagonal elements of the tridiagonal Hamiltonian matrix obtained by the Lanczos method with initial vector $\hat{S}_{\mathbf{q}}^{\alpha}|\psi_0 \rangle$. Here, we show the ED results using $24b$ cluster in Fig.~\ref{fig:AppQSLSqw}. At $J_1^{\pm\pm}=-0.17, J_1^{z\pm}=0.6, J_2=0.0$, though there are finite-size effects, we still can observe that the low-energy maxima are located at M points. And the maxima at K points are at higher energy. It seems that our ED calculations are consistent with the inelastic neutron scattering measurements of YbMgGaO$_4$~\cite{Shen2016, Paddison2017}. While using the C set of parameters, the maxima in K and M points are nearly at the same energy. It can be easy to understand this phenomenon. Starting from 120$^{\circ}$ N\'eel phase to the QSL phase, and then to a stripe phase, the spectral weight would transfer from K points to the M points.

\section{Exchange parameters}

In Table~\ref{table:AppParam}, we list three sets of exchange parameters fitted by experimental data and got from Ref.~\onlinecite{Paddison2017}, Ref.~\onlinecite{YDLi2018} and Ref.~\onlinecite{Haravifard2020}, respectively. $A$ set of parameters was used to calculate the specific heat in Fig.~\ref{fig:AppExperimentCv}(b). $C$ set of parameters was used to calculate the magnetization curves in Figs.~\ref{fig:AppMagFieldII}(b1) and \ref{fig:AppMagFieldII}(b2). $B$ set of parameters was used to calculate the specific heat in Fig.~\ref{fig:AppExperimentCv}(a) and Fig.~\ref{fig:GlassOrder}(c) of the main text. In Appendix, we also use it to show the frustration parameter in Fig.~\ref{fig:AppFrusParam} of Appendix~\ref{App:Frustration},  the magnetization curves in Figs.~\ref{fig:AppMagFieldIII}(b1) and \ref{fig:AppMagFieldIII}(b2),  the square sublattice magnetization for Stripe-I phase in Fig.~\ref{fig:AppRMagOrdII}(d), the histograms of spin correlations in Figs.~\ref{fig:AppQSLRHistogram}(c) and \ref{fig:AppQSLRHistogram}(d),  the distribution of spin domains with different number of spins changing with $\Delta$ in Fig.~\ref{fig:AppQSLRHistogramII}(b) and the nearest-neighbor spin correlations for different random configurations in Figs.~\ref{fig:BNNCorr}(b1)--\ref{fig:BNNCorr}(b4) of the main text.

\begin{table}[htp!]
  \centering
  \caption{\label{table:Exchange Parameter}Three sets of exchange parameters get from Ref.~\onlinecite{Paddison2017}, Ref.~\onlinecite{YDLi2018}, and Ref.~\onlinecite{Haravifard2020}, respectively.}
  \begin{tabularx}{0.4\textwidth}{ccccc}
  \hline\hline
  \quad          \qquad\qquad &$A$   \qquad\qquad &$B$     \qquad\qquad &$C$ \\
  \hline
  $J\ (\rm{meV})$       \qquad\qquad &0.126 \qquad\qquad &0.164   \qquad\qquad &0.164 \\
  $J_1^{zz}$     \qquad\qquad &1     \qquad\qquad &1       \qquad\qquad &1 \\
  $J_1^{\pm}$    \qquad\qquad &1.73  \qquad\qquad &1.317   \qquad\qquad &1.317 \\
  $J_1^{\pm\pm}$ \qquad\qquad &0.103 \qquad\qquad &0.341   \qquad\qquad &0 \\
  $J_1^{z\pm}$   \qquad\qquad &0     \qquad\qquad &0.598 \qquad\qquad &0.13 \\
  $J_2/J_1$      \qquad\qquad &0.22  \qquad\qquad &0       \qquad\qquad &0.1 \\
  $J_2^{zz}$     \qquad\qquad &0.22  \qquad\qquad &0       \qquad\qquad &0.1 \\
  $J_2^{\pm}$    \qquad\qquad &0.381 \qquad\qquad &0       \qquad\qquad &0.132 \\
  \hline\hline
  \end{tabularx}
  \label{table:AppParam}
\end{table}

\bibliography{YMGO}

\end{document}